\documentclass[12pt,preprint]{aastex}
\usepackage[numberedappendix]{emulateapj5}
\usepackage{amstext}
\usepackage{apjfonts}
\usepackage{amsmath}

\newcommand{\nablas}{{\mathbf{\nabla}}\!\!_s}
\newcommand{\mbff}{{\mathbf {f}}}
\newcommand{\mbffone}{{\mathbf {f}}}

\begin{document}
\submitted {Submitted to {\apj} on March 19, 2003}
\journalinfo {Submitted to {\apj} on March 19, 2003}

\shorttitle{RELATIVISTIC MHD AND GRB OUTFLOWS: I.}
\shortauthors{{VLAHAKIS} AND {K\"ONIGL}}

\title{
Relativistic Magnetohydrodynamics with Application to Gamma-Ray Burst Outflows:
I. Theory and Semianalytic Trans-Alfv\'enic Solutions}

\author{Nektarios Vlahakis and Arieh K\"onigl}
\affil{Department of Astronomy \& Astrophysics and Enrico Fermi
Institute, University of Chicago, 5640 S. Ellis Ave., Chicago, IL 60637}
\email{vlahakis@jets.uchicago.edu, arieh@jets.uchicago.edu}

\begin{abstract}
We present a general formulation of special-relativistic
magnetohydrodynamics and derive exact radially self-similar
solutions for axisymmetric outflows from strongly magnetized, rotating compact
objects. We generalize previous work by including thermal
effects and analyze in detail the various forces that
guide, accelerate, and collimate the flow. We demonstrate that,
under the assumptions of a quasi-steady poloidal magnetic field
and of a highly relativistic poloidal velocity, 
the equations become
effectively time-independent and the motion can be described as
a frozen pulse. We concentrate on
trans-Alfv\'enic solutions and consider outflows that are
super-Alfv\'enic throughout in the companion paper. Our results
are applicable to relativistic jets in gamma-ray burst (GRB)
sources, active galactic nuclei, and microquasars, but our
discussion focuses on GRBs. We envision the outflows in this
case to initially consist of a hot and optically thick mixture of
baryons, electron-positron pairs, and photons. We show that the
flow is at first accelerated thermally but that the bulk of the acceleration is
magnetic, with the asymptotic Lorentz factor corresponding to a
rough equipartition between the Poynting and kinetic-energy
fluxes (i.e., $\sim 50 \%$ of the injected total energy
is converted into baryonic kinetic energy). The electromagnetic
forces also strongly collimate the flow, giving rise to an
asymptotically cylindrical structure.
\end{abstract}


\section{Introduction}
\label{introduction}

The acceleration and collimation of powerful bipolar outflows
and jets in a variety of astronomical settings are often
attributed to the action of magnetic fields (see, e.g.,
\citealt{L00} and \citealt{KPu00} for reviews). The commonly
invoked scenario is that magnetic field
lines threading a rotating compact object or its
surrounding accretion disk can efficiently tap the rotational
energy of the source and accelerate gas to supersonic speeds through
centrifugal and/or magnetic pressure-gradient forces. It is
argued that the hoop stresses of the twisted field lines can
account for the narrowness of many jets and that, in many cases, alternative
production mechanisms (such as thermal driving) can be excluded
on observational grounds.

Although numerical simulations have provided useful
insights into various aspects of hydromagnetic jet production,
practical limitations have necessitated complementing this
approach with analytic studies. Owing to the complexity of
the problem, the most general semianalytic solutions obtained so far have
been time-independent and self-similar, patterned after the pioneering
disk-outflow solutions of \citet{BP82}. The advantage of pursuing
such solutions is that they are exact and
self-consistent, and that they can be systematically classified
\citep[e.g.,][]{VT98}. Furthermore, these solutions
are evidently rich enough to capture most of the relevant physics,
as corroborated by numerical calculations \citep[e.g.,][]{OP97,U99,KLB99}.

Almost all of the previous semianalytic work on jet
magnetohydrodynamics (MHD) was done in the Newtonian
limit of nonrelativistic bulk and random speeds. Relativistic
outflows are, however, observed quite commonly in Nature ---
with examples including active galactic nuclei (AGNs), Galactic
superluminal sources (often referred to as ``microquasars''),
pulsars, and gamma-ray bursts
(GRBs)--- and in many of these cases magnetic fields are again
implicated as the main driving and collimation mechanism
\citep[e.g.,][]{B02a}. This provides a motivation
for generalizing the Newtonian self-similar outflow
solutions, although it is readily seen that this cannot be done
in a totally straightforward manner. For one thing,
special-relativistic MHD (unlike the nonrelativistic theory)
involves a characteristic speed (the speed of light $c$), which
precludes the incorporation of gravity into the self-similar
equations and a simple matching of the outflow solution to a particular
(e.g., Keplerian) disk rotation law. Furthermore, again in
contrast with the nonrelativistic formulation, the displacement
current and the charge density cannot be neglected in the
constitutive equations (which now must also satisfy relativistic
covariance).

Despite the aforementioned complications, \citet*{LCB92} and
\citet{C94} succeeded in generalizing the ``cold,'' radially
self-similar solutions of \citet{BP82} to the relativistic regime. 
Their solutions are
characterized by the thermal pressure playing a negligible role
in the flow acceleration and by 
the flow being trans-Alfv\'enic: the poloidal speed is less
than the poloidal component of the Alfv\'en speed at the base of
the flow, and comes to exceed it further up.
Our aim in this
paper is to further generalize these solutions to the ``hot''
relativistic case --- i.e., we allow both the bulk {\em
and} the random speeds to be relativistic. This is motivated
primarily by the desire to apply these solutions to GRB
outflows, in which thermal driving by an optically thick, hot ``fireball''
composed primarily of radiation and electron-positron pairs
could play a role. In fact, most previous models of GRB outflows
were purely hydrodynamical and included only thermal driving
(see, e.g., \citealt{P99} for a review). It
has subsequently been realized that energy deposition by
annihilating neutrinos, which had been one of the main proposed
heating mechanisms, would typically be inefficient \citep[e.g.,][]{DPN02},
and it was, furthermore, argued
\citep{DM02} that only a small fraction ($\lesssim 1\%$) of
the energy deposited in the source could be in thermal form to
avoid generating strong photospheric emission in the outflow (for
which there has been no observational evidence). For these (and
other) reasons, it is now
believed that magnetic fields play the dominant role in the
driving of GRB jets (e.g., \citealt{M02}). Nevertheless,
some thermal energy injection, either by neutrinos or by
magnetic energy dissipation, is likely to take place
\citep*[e.g.,][]{NPP92,MLR93}, and, as we show in \S \ref{soln_a}, even if it
contributes only a small fraction of the initial energy flux in
the flow, it can dominate the early phase of the
acceleration. There are indications that thermal energy
deposition at the source may also contribute to the initial acceleration of
relativistic jets in AGNs (e.g., \citealt*{MLF02}; N. Vlahakis \&
A. K\"onigl, in preparation). It thus appears that, to fully
understand the nature of such jets, it is necessary to model them in
terms of a ``hot'' MHD outflow.

The ability of large-scale, ordered magnetic fields to guide,
collimate, and accelerate relativistic outflows has been previously
discussed by various authors (including, in the GRB
context, \citealt{U94},
\citealt{T94}; \citealt{MR97}, \citealt{K97}, \citealt{KR98},
\citealt{LB01}, and \citealt{DS02}). These discussions, however,
did not include
exact solutions from which detailed quantitative estimates could
be made. In a previous paper, Vlahakis \& K\"onigl (2001,
hereafter VK01) presented a semianalytic self-similar
solution of the ``hot'' relativistic MHD equations and applied
it to the interpretation of GRBs. The full formalism underlying
this solution is described in the present paper, where we also
analyze its dependence on the relevant physical parameters and
compare it with other characteristic solutions of trans-Alfv\'enic flows.
This discussion is extended in the companion paper (Vlahakis \&
K\"onigl 2003, hereafter Paper II), where we focus on super-Alfv\'enic flows.

The self-similar solutions that we derive correspond to ordered
magnetic field configurations in the ideal-MHD limit. Although
it is quite conceivable that the fields that drive the flow from
a differentially rotating star or disk are at least in part
small-scale and disordered, the statistical (temporal and
spatial) averages of such fields could in principle have a
similar effect to that of large-scale, ordered field
configurations in providing both acceleration \citep{HB00} and
collimation \citep{L02}. The applicability of
ideal MHD to the acceleration of ultrarelativistic flows has
been questioned by \citeauthor{B02b} (\citeyear{B02b}; see also
\citealt{LB02}), who proposed instead a force-free
electromagnetic formulation. We note in this connection that a
force-free behavior can be recovered from the relativistic MHD
formulation as a limiting case of negligible particle
inertia. Furthermore, electromagnetic energy dissipation --- perhaps
the strongest argument against the ideal-MHD modeling framework
--- has been claimed to lead, on its own, to an efficient
conversion of Poynting flux into a highly relativistic bulk motion
\citep[e.g.,][]{DS02}.
Our ideal-MHD solutions may thus be regarded as a first step
toward a more comprehensive theory in which dissipation effects
will be taken into account (and possibly further enhance the
effectiveness of the magnetic acceleration process).

Although the formulation presented in this paper is quite
general, the main application that we consider is to GRB
sources. For definiteness, we adopt the ``internal shock''
scenario for the origin of the prompt high-energy emission in
GRBs \citep[e.g.,][]{P99}. Considerations involving
variability time scales (as interpreted in the context of this picture) as well as
source energetics support the identification of an accretion
disk around a newly formed black hole as the source of the GRB
outflow \citep[e.g.,][]{P01a}; accordingly, we
concentrate on modeling jets from accretion disks. However, our
solutions should be at least qualitatively applicable also to
other source configurations, such as a rapidly spinning neutron
star \citep[e.g.,][]{U94,KR98}
or a rotating, magnetically threaded black hole \citep[e.g.,][]{BZ77,V01}.
Since GRB outflows have a
limited duration (the value of which is
plausibly related to the disk accretion time), a naive
application of a steady-state similarity solution is not
warranted. In previous, purely hydrodynamical models of GRB outflows,
this difficulty was circumvented by applying the so-called ``frozen pulse''
approximation \citep*{PSN93}. In this paper we prove (\S
\ref{governing}) that this approximation can be generalized to the
relativistic MHD regime, but we also demonstrate (\S \ref{timerecovery})
how any inherent time dependence can be recovered.

The plan of the paper is as follows. 
In \S \ref{MHD} we present the equations of time-dependent
relativistic MHD, simplify them using the frozen-pulse approximation, and discuss
what effect each of the various forces acting on the plasma has
on the flow acceleration and collimation.
In \S \ref{r-ss_section} we describe the $r$ self-similar model
and in \S \ref{results} we give the results of the numerical
integration of the model equations. We discuss general
implications of this work to GRB sources and other relativistic
jet sources in \S \ref{conclusions}, where we also summarize our conclusions.

\section{The Relativistic MHD Formulation}\label{MHD}

\subsection{Governing Equations}\label{governing}

The stress-energy tensor of relativistic MHD consists of three
parts --- matter (subscript M), radiation (subscript R) and
electromagnetic fields (subscript EM): $ T^{\kappa \nu}=T^{\kappa \nu}_{\rm M}
+T^{\kappa \nu}_{\rm R}+T^{\kappa \nu}_{\rm EM}$ ($\kappa\,,\nu = 0,1,2,3$). The matter
component is given by $T^{\kappa \nu}_{\rm M}=\left(\rho_{\rm M} + 
P_{\rm M} /c^2 \right) U^{\kappa} U^{\nu} + P_{\rm M} \eta^{\kappa \nu}$,
where $\rho_{\rm M} c^2=\rho_0 c^2+\rho_0 e_{\rm M}$ is the total
comoving matter energy density, $P_{\rm M}$ is the particle
pressure, $U^{\nu}=\left(\gamma c,\gamma \boldsymbol{V}\right)$
is the fluid four-velocity, and $\eta^{\kappa \nu}={\rm
diag} \left(-1\,1\,1\,1\right)$ is the metric tensor (assuming a
flat spacetime and Cartesian space coordinates $x_j,\ j=1,2,3$). 
Here $\rho_0$ is the baryon rest-mass density,
$\rho_0 e_{\rm M}=P_{\rm M}/(\Gamma-1)$ is the internal energy density, with
$\Gamma$ denoting the polytropic index ($=4/3$ or $5/3$ in the
limit of an ultrarelativistic or a nonrelativistic temperature, respectively),
$\boldsymbol{V}$ is the three-velocity measured in the frame of
the central object, and $\gamma=1/(1-V^2/c^2)^{1/2}$ is the Lorentz
factor. The radiation component is given by
$T^{\kappa \nu}_{\rm R}=\left(\rho_{\rm R} + 
P_{\rm R} /c^2 \right) U^{\kappa} U^{\nu} + P_{\rm R} \eta^{\kappa \nu}$,
where $\rho_{\rm R} c^2$ and $P_{\rm R}$ are, respectively, the
radiation energy density and pressure in the fluid rest frame. Radiation forces are
typically most important in regions that are sufficiently
optically thick that one can take the local radiation field to
be nearly isotropic and set $\rho_{\rm R}c^2=3 P_{\rm R}$.
We are most interested in the ultrarelativistic case
$\Gamma=4/3$, in which the matter and radiation can be treated
(under optically thick conditions)
as a single fluid. Thus, we henceforth write $\rho = \rho_{\rm M}+\rho_{\rm
R}$, $P=P_{\rm M}+P_{\rm R}$, and $\rho_0 e= \rho_0 e_{\rm M}+\rho_{\rm R} c^2$.

Introducing the specific (per baryon mass) relativistic enthalpy
$\xi c^2$, where
\begin{equation}\label{xi}
\xi = \frac{\rho c^2 + P}{\rho_0 c^2} 
=1+\frac{\Gamma}{\Gamma-1}\frac{P}{\rho_0 c^2}
\end{equation}
and including the contribution of the electric
($\boldsymbol{E}$) and magnetic ($\boldsymbol{B}$) fields
(measured in the central-object frame), the components of the total
stress-energy tensor take the form ($j\,,k = 1,2,3$)
\begin{mathletters}
\begin{eqnarray}
T^{00}&=&\gamma^2 \xi \rho_0 c^2 -P + \frac{E^2+B^2}{8 \pi}
\,, \\
T^{0j}&=&T^{j0}=\left( \xi \rho_0 c \gamma^2 {\boldsymbol V}+
\frac{{\boldsymbol E} \times {\boldsymbol B}}{4 \pi} \right) \cdot \hat{x}_j 
\,, \label{tensor_flux}\\
T^{jk}&=&\xi \rho_0 \gamma^2 {V_j V_k }-
\frac{E_j E_k + B_j B_k}{4 \pi} 
+\left(P+ \frac{ E^2 + B^2}{8 \pi}
\right) \eta^{jk} \, , 
\end{eqnarray}
\end{mathletters}
with $T^{00}$, $c T^{0j} \hat{x}_j$, and $T^{jk}$ representing
the energy density, energy flux, and spatial stress
contributions, respectively.

The electromagnetic field obeys Maxwell's equations
\begin{eqnarray} 
{\boldsymbol{\nabla}} \cdot {\boldsymbol{B}}=0 \,, \quad &&
{\boldsymbol{\nabla}} \cdot {\boldsymbol{E}}= \frac{4 \pi}{c} J^0 \,,
\nonumber
\\
{\boldsymbol{\nabla}} \times {\boldsymbol{B}}=\frac{1}{c}
\frac{\partial {\boldsymbol{E}}}{\partial t}+
\frac{4 \pi}{c}{\boldsymbol{J}}\, , \quad &&
{\boldsymbol{\nabla}} \times {\boldsymbol{E}}=-\frac{1}{c}
\frac{\partial {\boldsymbol{B}}}{\partial t}\ ,
\label{maxwell}
\end{eqnarray}
\noindent
where $J^{\nu}=\left(J^0,\boldsymbol{J}\right)$ is the
four-current (with $J^0/c$ representing the charge
density). Under the assumption of ideal MHD, the
comoving electric field is zero, which implies
\begin{equation}\label{ohm}
{\boldsymbol{E}}=-\frac{\boldsymbol{V}}{c} \times \boldsymbol{B}\ .
\end{equation}

The baryon mass conservation equation is $(\rho_0
U^{\nu})_{\,,\nu}=0$, or
\begin{equation}\label{continuity}
\left( \frac{\partial}{\partial t} + {\boldsymbol{V}} \cdot
{\boldsymbol{\nabla}} \right) \left(\gamma \rho_0 \right) +
\gamma \rho_0 {\boldsymbol{\nabla}} \cdot {\boldsymbol{V}} =0 \, .
\end{equation}
In the absence of a gravitational field or any other
external force, the equations of motion are $T^{\kappa \nu}_{\,,\nu}=0$.
The momentum conservation equation is given by the $\kappa=1,2,3$ components,
\begin{equation}\label{momentum}
\gamma \rho_0
\left( \frac{\partial}{\partial t} + {\boldsymbol{V}} \cdot
{\boldsymbol{\nabla}} \right)
\left(\xi \gamma {\boldsymbol{V}} \right)=
-{\boldsymbol{\nabla}}P +\frac{J^0 {\boldsymbol{E}}
+{\boldsymbol{J}} \times {\boldsymbol{B}} }{c}\ .
\end{equation}
The entropy conservation equation (the first law of thermodynamics)
is obtained by setting $U_{\kappa} T^{\kappa \nu}_{\,,\nu}=0$,
\begin{eqnarray}
\left( \frac{\partial}{\partial t} + {\boldsymbol{V}} \cdot
{\boldsymbol{\nabla}} \right)e+ P
\left( \frac{\partial}{\partial t} + {\boldsymbol{V}} \cdot
{\boldsymbol{\nabla}} \right) \left( \frac{1}{\rho_0}\right) =0\ ,
\nonumber
\end{eqnarray}
which can be rewritten (using $\rho_0
e={P}/\left[{\Gamma-1}\right]$) as
\begin{equation}\label{energy}
\left( \frac{\partial}{\partial t} + {\boldsymbol{V}} \cdot
{\boldsymbol{\nabla}} \right) \left(\frac{P}{\rho_0^\Gamma}\right)=0\,.
\end{equation}

One can carry out a partial integration of equations
(\ref{maxwell})--(\ref{energy}) under the assumptions of
axisymmetry [in cylindrical coordinates ($\varpi\,,\phi\,,z$),
$\partial / \partial \phi=0$] and of a zero azimuthal electric
field ($E_{\phi}=0$) if the flow is time-independent \citep[e.g.,][]{
BO78,L86}. We now
show that the equations describing a highly relativistic MHD ``pulse'' that could be
identified with a GRB event may, in fact, be cast in a steady-state form. 
We start by noting that, with the above assumptions
$\left( \partial / \partial \phi=0 \,, E_{\phi}=0 \right)$,
the poloidal component of Faraday's law implies that the poloidal
magnetic field is time independent.
If the field is anchored in an accretion disk, then the poloidal
field near the disk surface would remain quasi steady 
at least on the timescale of the local radial inflow ($\sim
\varpi/|V_\varpi|$). In the case of a GRB outflow
associated with the emptying up (by accretion onto a black hole)
of a disk of finite size, the poloidal field may be
expected to change significantly only on the timescale of the
burst duration. At the end of this time interval, the information that the field
has changed starts to propagate with at most the speed
of light (the actual speed of propagation is the fast
magnetosonic speed, which is generally $< c$ in a material
medium). As we are concerned with highly relativistic outflows, it is reasonable
to expect that the poloidal field associated with the outflowing
fluid elements that produce the burst will exhibit negligible explicit
time dependence ($\partial/\partial t\approx 0$) over the duration of the burst.
(Note, however, that the azimuthal component of the magnetic field,
which is related to the Poynting flux, will be time dependent.)

The solenoidal condition on the magnetic field, ${\boldsymbol{\nabla}} \cdot
{\boldsymbol{B}} =0$, implies that there is a poloidal
magnetic flux function $A(\varpi\,, z)$, defined by $2 \pi A =
\int \! \! \! \! \int {\boldsymbol{B}}_p \cdot d {\boldsymbol
{S}}$, which satisfies 
\begin{equation}\label{solenoidal}
{\boldsymbol{B}}={\boldsymbol{B}}_p + {\boldsymbol{B}}_{\phi}\,, \quad
{\boldsymbol{B}}_p= \frac{{\boldsymbol{\nabla}}A \times \hat{\phi} }{\varpi}\ ,
\end{equation}
where the subscripts $p$ and $\phi$ denote the poloidal and
azimuthal components, respectively. Furthermore, equation
(\ref{ohm}) together with the condition $E_\phi=0$ implies
${\boldsymbol{V}}_p \parallel {\boldsymbol{B}}_p$, from which it
follows that there are functions $\Psi_A$ and $\Omega$ (whose
coordinate dependence we discuss below) such that
\begin{equation}\label{V-B}
{\boldsymbol{V}}=\frac{\Psi_A}{4 \pi \gamma \rho_0}
{\boldsymbol{B}} + \varpi \Omega \hat {\phi}
\,, \quad \frac{\Psi_A}{4 \pi \gamma \rho_0} = \frac{V_p}{B_p}
\ .
\end{equation}

Denoting the arclength along a poloidal fieldline by
$\ell(\varpi\,,z)$, we change variables from
($\varpi\,,z\,,t$) to ($A\,, \ell \,,s$), with $s \equiv ct-\ell$.
For any function $\Phi=\Phi(A\,, \ell \,,s)$, we can define the operator
\begin{equation}
\nablas \Phi \equiv 
\frac{\partial \Phi}{\partial A} {\boldsymbol{\nabla}} A +
\frac{\partial \Phi}{\partial \ell} {\boldsymbol{\nabla}} \ell =
{\boldsymbol{\nabla}} \Phi + \frac{\partial \Phi}{\partial s}
{\boldsymbol{\nabla}} \ell\ .
\end{equation}
We now rewrite the MHD equations using
\begin{equation}\label{operators}
{\boldsymbol{\nabla}} = \nablas -
{\boldsymbol{\nabla}}  \ell \ \frac{\partial}{\partial s} \ ,
\quad \frac{\partial}{\partial t}=c \frac{\partial}{\partial s}\ .
\end{equation}

\begin{mathletters}\label{MHD-eqs-s}
Equation (\ref{solenoidal}) becomes
\begin{equation}
{\boldsymbol{B}}= \frac{\nablas A \times
\hat{\phi} }{\varpi} + {\boldsymbol{B}}_{\phi}\,,
\end{equation}
whereas equation (\ref{ohm}) gives
\begin{equation}\label{E_A}
{\boldsymbol{E}}=-\frac{\Omega}{c} \nablas A
\,, \quad E=\frac{\varpi \Omega}{c} B_p\,.
\end{equation}
Faraday's law implies
\begin{equation}\label{Omega_A}
\nablas \Omega  \times \nablas A =
c \frac{\partial (E+B_{\phi})}{\partial s}\hat{\phi} \,,
\end{equation}
and the continuity equation yields
\begin{equation}\label{continuity-s}
\nablas \cdot (4 \pi \gamma \rho_0 {\boldsymbol{V}})+
(c-V_p) \frac{\partial (4 \pi \gamma \rho_0)}{\partial s}-
4 \pi \gamma \rho_0 \frac{\partial V_p }{\partial s}=0\ ,
\end{equation}
where, using equation (\ref{V-B}), $\nablas
\cdot (4 \pi \gamma \rho_0 {\boldsymbol{V}})=
\left( \nablas \Psi_A \times
\nablas A \right) \cdot \hat{\phi} / \varpi$.

Turning now to the momentum conservation equation, we employ equation
(\ref{operators}) and the fact that ${\boldsymbol{B}}_p$ is time independent 
$\left(\partial {\boldsymbol{B}}_p / \partial s =0\right)$ to
write the current density in the form
$${\boldsymbol{J}} = 
\frac{c}{4 \pi} \left[ \nablas \times {\boldsymbol{B}}
+ \frac{\partial}{\partial s} \left(
{\boldsymbol{B}}_{\phi} \times \nablas \ell -{\boldsymbol{E}} 
\right) \right] \ . $$
Decomposing the vectors using the local Cartesian basis
$$\left(\hat{n}\equiv 
\frac{{\boldsymbol E}}{E}=
-\frac{\nablas A}{
\mid \nablas A \mid } \,, \
\hat{b}\equiv \frac{{\boldsymbol B}_p}{B_p} \,, \ \hat{\phi} \right)\ ,$$
we get
\begin{eqnarray}
\nonumber
{\boldsymbol{B}}_{\phi} \times \nablas \ell -{\boldsymbol{E}} &=&
B_{\phi} \hat{\phi} \times \left[
\left(\hat{b} \cdot \nablas \ell \right) \hat{b}+
\left(\hat{n} \cdot \nablas \ell \right) \hat{n} 
\right] - E \hat{n} 
\\ \nonumber
&=&
-\left(E+ B_{\phi} \right) \hat{n}
+B_{\phi} \left(\hat{n} \cdot \nablas \ell \right) \hat{b} 
\,,
\end{eqnarray}
and hence
$$ 
{\boldsymbol{J}}= \frac{c}{4 \pi}  \left[
\nablas \times {\boldsymbol{B}}
- \hat{n} \frac{\partial \left(E + B_{\phi}\right) }{\partial s} 
+\hat{b} \left(\hat{n} \cdot \nablas \ell \right)
\frac{\partial B_{\phi}}{\partial s} \right]\ .
$$
The charge density, in turn, can be written as
$$\frac{1}{c} J^0=\frac{1}{4 \pi}
{\boldsymbol{\nabla}}\cdot
\left( E \hat{n} \right)=
\frac{1}{4 \pi} \left[ 
\nablas {\boldsymbol{E}} -
\left(\hat{n} \cdot \nablas \ell  \right)
\frac{\partial E }{\partial s}
\right]\ .
$$
Substituting these expressions into equation (\ref{momentum}),
we obtain
\begin{eqnarray}\label{momentum-s}
& &
\gamma \rho_0 \left({\boldsymbol{V}} \cdot \nablas \right)
\left(\xi \gamma {\boldsymbol{V}} \right) + \gamma \rho_0 (c-V_p)
\frac{\partial}{\partial s} \left(\xi \gamma {\boldsymbol{V}} \right)=
\nonumber \\ & &
\frac{\partial (E+B_{\phi})}{4 \pi \partial s}
\frac{\nablas A}{\mid \nablas A
\mid} \times {\boldsymbol{B}}+\nablas A
\frac{\nablas \ell \cdot \nablas
A}{\mid \nablas A \mid ^2}
\frac{\partial (B_{\phi}^2-E^2)}{8 \pi \partial s}
\nonumber \\ & &
-{\boldsymbol{\nabla}}P
+\frac{\Omega}{4 \pi c^2} \left[\nablas 
\cdot
\left( \Omega \nablas A \right) \right]
\nablas A + \frac{\left(\nablas 
\times {\boldsymbol{B}} \right) \times {\boldsymbol{B}} } {4 \pi}
\ .
\end{eqnarray}
Finally, the entropy 
conservation equation transforms into
\begin{equation}\label{energy-s}
{\boldsymbol{V}} \cdot
\nablas \left(P / \rho_0^{\Gamma}\right) + (c-V_p) \frac{\partial 
\left(P/ \rho_0^{\Gamma}\right) }{\partial s}=0\ .
\end{equation}
\end{mathletters}

For a highly relativistic poloidal motion, when
$\gamma_p \equiv \left(1-V_p^2/c^2\right)^{-1/2} \approx \gamma \gg 1$,
one can simplify the equations by noting the following:
$\ $ 1) Due to Lorentz contraction, the observed width of the outflow is
$\gamma$ times smaller than its comoving width,
$\partial / \partial s \sim \gamma \partial /\partial \ell$.
As $c-V_p \approx c/ 2 \gamma^2$ and
${\boldsymbol{V}} \cdot {\boldsymbol{\nabla}} = V_p \partial / \partial \ell$,
one gets $(c-V_p) \partial / \partial s \sim (1/ 2 \gamma)
{\boldsymbol{V}} \cdot {\boldsymbol{\nabla}}
\ll {\boldsymbol{V}} \cdot {\boldsymbol{\nabla}}$.
Thus, all terms in equations (\ref{MHD-eqs-s}) containing
$(c-V_p)$ are negligible.
$\ $ 2) The term $4 \pi \gamma \rho_0 \partial V_p / \partial s$ in
equation (\ref{continuity-s}) is of the order of
$(4 \pi \rho_0 / \gamma) {\boldsymbol{V}} \cdot
\nablas \gamma$ and is thus negligible in
comparison with the first term on the left-hand side of this equation.
The arguments above were originally given in
the context of a purely hydrodynamic (HD) flow by \citet{PSN93}.
$\ $ 3) Using 
$1-V_p/c \approx 1/2 \gamma_p^2$, $V_{\phi}/c < (1-V_p^2/c^2)^{1/2} = 1/\gamma_p$,
and equations 
(\ref{V-B}) and (\ref{E_A}), we infer
$-(E+ B_{\phi}) /B_p =(1-V_p/c) \varpi \Omega / V_p - V_{\phi}/V_p
\lesssim (\varpi \Omega/c)(1/2\gamma_p^2)
$, which
remains $\ll 1$ throughout the flow in view of the scaling $\gamma \propto \varpi$
(see \S~\ref{validity}).
Thus, all terms in equation (\ref{momentum-s}) that contain
$(E+B_{\phi})$ (i.e., the first two terms on the right-hand side) 
are a factor $\sim \gamma$ smaller than the 
last term on the right-hand side and can be neglected.
The same is true in equation (\ref{Omega_A}), which implies 
$\partial \Omega \approx \gamma c \partial [(E+B_\phi) / \varpi B_p]
\sim \gamma \partial (\Omega / \gamma^2) \ll \Omega$.
$\ $ 4) The pressure-gradient force in the momentum conservation
equation can also be neglected. It is much smaller than the
Lorentz force in the transfield direction, consistent with the
fact that the field is everywhere close to being force-free
in that direction
(especially so in the region near the origin, where thermal
effects are most important). Along the field, there is a force
that is $\sim \gamma^2$ times larger (namely, the inertial force
component associated with the ${\boldsymbol {V}}
\cdot {\boldsymbol{\nabla}} \xi$ term), as in the
purely HD case examined in \citet{PSN93}.
In general, the pressure force is important up to the slow magnetosonic point, where,
for highly relativistic temperatures, $\gamma \sim (3/2)^{1/2}$, and
its contribution is negligible in the highly relativistic
regime. One can therefore replace the term $-{\boldsymbol{\nabla}} P$ by
$-\nablas P$ or even completely neglect the
pressure-gradient force in the momentum equation without introducing a significant
error.

With the above approximations, all the $\partial/\partial s$
terms in the continuity, momentum, and entropy equations can be
eliminated, and the conservation equations simplify to a
steady-state form. Although the label $s$ remains attached to
the $\nabla$ operator, it now serves only to identify a given
outflowing shell (or pulse). The motion remains effectively time independent
and can be described as a frozen pulse whose internal
profile is specified through the variable $s$. As we noted
in \S \ref{introduction}, the frozen-pulse approximation was first applied to
relativistic HD outflows in GRB sources by \citeauthor{PSN93}
(\citeyear{PSN93}; see also \citealt{P99}). We have now shown
that this approximation can be extended to relativistic MHD flows. In the
remainder of this paper we pursue this effectively
steady-state formulation, but we return in \S \ref{timerecovery}
to consider time-dependent effects in GRB outflows.

The full set of effectively steady-steady equations can
be partially integrated to yield several field-line constants: 
\smallskip
\begin{mathletters}\label{integrals-s}

\noindent a) Equations (\ref{V-B}) and (\ref{Omega_A}) yield the field angular
velocity, which equals the matter angular velocity at the
footpoint of the fieldline at the midplane of the disk,
\begin{equation}\label{int-Omega}
\Omega = \Omega(A\,, s)=
\frac{V_{\phi}}{\varpi}-\frac{\Psi_A}{4 \pi \gamma \rho_0}\frac{B_\phi}{\varpi}\ .
\end{equation}

\noindent b) The continuity equation (\ref{continuity-s}) and
equation (\ref{V-B}) imply that the mass-to-magnetic flux
ratio has the form
\begin{equation}
\Psi_A=\Psi_A(A\,, s)= \frac{4 \pi \gamma \rho_0 V_p}{B_p}\ .
\end{equation}

\noindent c) The $\hat{\phi}$ component of the momentum equation
(\ref{momentum-s}) yields the total (kinetic + magnetic) specific angular momentum,
\begin{equation}\label{int-L}
L=L(A\,, s) = \xi \gamma \varpi V_\phi -\frac{\varpi B_\phi}{\Psi_A}\ .
\end{equation}

\noindent d) Dotting ${\boldsymbol{V}}$ into the momentum equation
(\ref{momentum-s}) gives the total energy-to-mass flux ratio $\mu c^2$, where
\begin{equation}\label{int-mu}
\mu=\mu(A\,, s)= \xi \gamma - \frac{\varpi \Omega B_\phi}{\Psi_A c^2}\ .
\end{equation}

\noindent e) The entropy equation (\ref{energy-s}) gives the adiabat
\begin{equation}
\label{polytropic}
Q=Q(A\,, s)= \frac{P}{\rho_0^{\Gamma}}\ .
\end{equation}
Equation (\ref{polytropic})  is the usual polytropic relation between density and
pressure, but in the current application the polytropic index is
only allowed to take the values $4/3$ (if the temperatures are
relativistic, in which case matter and radiation are treated as
a single fluid) and $5/3$ (if the gas is ``cold,'' in which case
radiation forces can be neglected).\footnote{Any value of $\Gamma$
other than 4/3 or 5/3
would imply a nonadiabatic evolution and
hence require the incorporation of heating/cooling terms into
the entropy and momentum equations.}
\end{mathletters}

Two integrals remain to be performed, involving the Bernoulli and
transfield equations. There are correspondingly two unknown functions,
which we choose to be 
the cylindrical radius of the fieldline in units of the ``light cylinder'' radius,
\begin{equation}
x\equiv \varpi \Omega / c \,, 
\end{equation}
and the ``Alfv\'enic'' Mach number \citep[see][]{M69}
\begin{equation}\label{Mach}
M\equiv ( \gamma V_p / B_p)(4 \pi \rho_0 \xi)^{1/2}=\Psi_A (\xi
/ 4 \pi \rho_0)^{1/2}\,.
\end{equation}

We define the Alfv\'en lever arm by $\varpi_A \equiv (L/
\mu \Omega)^{1/2}$ [and correspondingly $x_A \equiv \varpi_A \Omega /c
= (L \Omega/ \mu c^2)^{1/2}$] and use it to scale the
cylindrical radius of the fieldline by introducing
\begin{equation}\label{G}
G \equiv \varpi/\varpi_A =x/x_A\,.
\end{equation}
To obtain nondimensional variables, we adopt a reference length $\varpi_0$ and
a reference magnetic field $B_0$ and define
\begin{equation}\label{alpha}
\alpha \equiv \frac{\varpi_A^2}{\varpi_0^2}= \frac{\varpi^2}{\varpi_0^2 G^2}\,,
\quad \ \ {\cal A}\equiv \frac{2}{B_0 \varpi_0^2}A \ .
\end{equation}

The expressions for the physical quantities in terms of the
defined variables and the explicit expressions for the Bernoulli
and transfield equations are given in Appendix \ref{RMHDeqs}.
Except for the $s$ label, which serves to identify a given shell
(or pulse), these equations are precisely those of steady-state,
relativistic MHD. Solving these equations requires the
specification of seven constraints, of which
four are associated with boundary conditions at the source and
three are determined by the regularity requirement at the singular points
(the modified-slow, Alfv\'en, and modified-fast points; see \citealt{V00}).

\subsection{Forces in the Poloidal Plane}\label{forces}

The momentum equation (\ref{momentum}) can be written as the sum of the following
force densities (for simplicity we use hereafter the term force)
\begin{equation}\label{momentum-f}
{\mbff}_G + {\mbff}_T + {\mbff}_C +
{\mbff}_I + {\mbff}_P+{\mbff}_E+{\mbff}_B=0 \,,
\end{equation}
where
\begin{mathletters}
\begin{eqnarray}
&& \left.
\begin{array}{lcl}
{\mbff}_G = -\gamma \rho_0 \xi
\left({\boldsymbol{V}}\cdot \nablas \gamma \right) {\boldsymbol{V}}
& & 
\\
{\mbff}_T = -\gamma^2 \rho_0 \left({\boldsymbol{V}}\cdot
\nablas \xi \right) {\boldsymbol{V}}
\ & : & \mbox{temperature force}
\\
{\mbff}_C =\hat{\varpi} \gamma^2 \rho_0 \xi V_{\phi}^2/ \varpi
\ & : & \mbox{centrifugal force}
\\
{\mbff}_I = -\gamma^2 \rho_0 \xi
\left({\boldsymbol{V}}\cdot \nablas \right)
{\boldsymbol{V}} - {\mbff}_C
& &
\end{array} \right\} 
\begin{array}{c} \mbox{inertial}\\ \mbox{force}\end{array}
\nonumber \\
&& \left. 
\begin{array}{lcl}
{\mbff}_P = -\nablas P
& : & \mbox{pressure force}
\\
{\mbff}_E = \left(\nablas \cdot
{\boldsymbol{E}}\right) {\boldsymbol{E}} / 4 \pi
& : & \mbox{electric force}
\\
{\mbff}_B = \left(\nablas \times
{\boldsymbol{B}}\right) \times {\boldsymbol{B}} / 4 \pi
\quad \ \ \
& : & \mbox{magnetic force}
\end{array} \right.
\nonumber 
\end{eqnarray}
\end{mathletters}

The ``gamma'' force ${\mbff}_G$ further decomposes into two terms:
${\mbff}_G = {\mbff}_{G_p}+ {\mbff}_{G_{\phi}}$, with
$$ {\mbff}_{G_p}=-\frac{\gamma^4 \rho_0 \xi}{2 c^2}
\left({\boldsymbol{V}}\cdot \nablas V_p^2 \right){\boldsymbol{V}} \,, \quad
{\mbff}_{G_{\phi}}=-\frac{\gamma^4 \rho_0 \xi}{2 c^2}
\left({\boldsymbol{V}}\cdot \nablas V_{\phi}^2 \right){\boldsymbol{V}} \,.$$
The poloidal part of the ${\mbff}_I$ force is 
$$
-\gamma^2 \rho_0 \xi \left({\boldsymbol{V}}_p\cdot
\nablas \right) {\boldsymbol{V}}_p 
= \gamma^2 \rho_0 \xi \left( V_p^2 \frac{\partial \psi}{\partial \ell}
\frac{\nablas A}{\mid \nablas A \mid} - 
\frac{\partial V_p}{\partial \ell} {\boldsymbol{V}}_p \right)  \,,$$
where 
$\psi$ is the angle between the poloidal magnetic field and the disk
($\cos \psi = V_{\varpi}/V_p$), and
the derivative $\partial / \partial \ell = \cos \psi \ \partial / \partial \varpi
=[\sin \theta \cos (\psi + \theta )/\varpi] \partial / \partial \theta) 
$ is taken keeping $A$ (and $s$) constant. The radius of
curvature of a poloidal fieldline is
${\cal R}= (\partial \psi / \partial \ell)^{-1}$.

\subsubsection{Poloidal Acceleration}\label{pol_accel}

The projection of equation (\ref{momentum-f}) along the poloidal flow is
\begin{eqnarray}\label{momentum-V_p}
&&
\frac{\gamma^2 \rho_0 \xi}{2}
\left(1+\gamma^2 \frac{V_p^2}{c^2} \right) \frac{\partial V_p^2}{\partial \ell}=
-\frac{\gamma^4 \rho_0 \xi V_p^2}{2 c^2} \frac{\partial V_{\phi}^2}{\partial \ell}
-\gamma^2 \rho_0 V_p^2 \frac{\partial \xi}{\partial \ell}
+ \nonumber \\ &&
+\gamma^2 \rho_0 \xi \frac{V_{\phi}^2}{\varpi} \cos \psi
-\rho_0 c^2 \frac{\partial \xi}{\partial \ell}
-\frac{B_{\phi}}{4 \pi \varpi} \frac{\partial \left(\varpi
B_{\phi} \right)}{\partial \ell}\ .
\end{eqnarray}
The terms on the right-hand side of equation
(\ref{momentum-V_p}) are recognized as
${\mbffone}_{G_{\phi} \parallel}$, ${\mbffone}_{T \parallel}$,
${\mbffone}_{C \parallel}$, ${\mbffone}_{P \parallel}$,
and ${\mbffone}_{B \parallel}\,,$ respectively,
where a subscript $\parallel$ denotes the component of a vector
along the poloidal fieldline. The first term on the left-hand side of equation
(\ref{momentum-V_p}) is $-{\mbffone}_{I \parallel}$,
whereas the second term is $-{\mbffone}_{G_p \parallel}$
(note that ${\mbffone}_{E \parallel}=0$).
The magnetic force component ${\mbffone}_{B \parallel}$ decomposes into the
azimuthal magnetic pressure gradient $-{\partial}
\left({B_{\phi}^2}/{8 \pi}\right)/{\partial \ell}$
and the magnetic tension $-{B_{\phi}^2} \cos \psi / {4 \pi \varpi}$.
These two parts cancel each other when $B_{\phi}^2(A\,,\varpi)\propto 1/ \varpi^2$;
if $B_{\phi}^2(A\,,\varpi)$ decreases faster than $\varpi^{-2}$
then the gradient of the azimuthal magnetic pressure exceeds the magnetic tension,
resulting in a positive ${\mbffone}_{B \parallel}$.

In the nonrelativistic
regime ($V \ll c\,, \ x \ll 1\,, \ \xi \approx 1$, ${\mbffone}_{G_{\phi} \parallel}=
{\mbffone}_{T \parallel}=0$), the
pressure force ${\mbffone}_{P \parallel}$
dominates up to the slow magnetosonic point, but
the bulk of the acceleration is either magnetocentrifugal ---
corresponding to the ${\mbffone}_{C \parallel}$ term,
which can be interpreted in the ``bead on a wire'' picture \citep[e.g.,][]{BP82}\footnote{
The strong poloidal magnetic fieldline plays the role of the wire.
In the cold limit one has $M \ll 1$ and $x\ll x_A$ (with $x_A\lesssim
1$) near the base of the flow, implying that the
azimuthal field satisfies $-\varpi B_{\phi} / L \Omega  \approx
1- G^2(1-x_A^2) + M^2 \approx 1$ 
and hence that the ${\mbffone}_{B \parallel}$ term is negligible
and that a near-corotation ($V_{\phi}/ \varpi \Omega \approx 1-M^2/G^2 \approx 1$) holds.
The small value of $M$ in turn implies a large density and hence
a measurable thermal pressure, resulting in a nonnegligible
pressure force at the base.} --- or a consequence of the
magnetic pressure-gradient force ${\mbffone}_{B \parallel}$
(which near the surface of the disk can be interpreted in the
``uncoiling spring'' picture; e.g., \citealt{US85}).\footnote{
In this picture, the winding-up of the fieldlines by the disk rotation produces a
large azimuthal magnetic field component that is antiparallel to $V_{\phi}$
in the northern hemisphere (and parallel to $V_{\phi}$ in the southern hemisphere),
and a corresponding outward-directed magnetic pressure gradient
$-\nablas (B_{\phi}^2/8 \pi )$.}

The magnetic force generally becomes important also in flows
where the centrifugal acceleration initially dominates: in this case
the inertia of the centrifugally accelerated gas amplifies the $B_\phi$ component, 
and eventually (beyond the Alfv\'en point)
${\mbffone}_{B \parallel}$ becomes the main driving
force. This force continues to accelerate the flow beyond the
classical fast-magnetosonic point (which separates the elliptic and hyperbolic regimes of the MHD
partial differential equations)\footnote{
At
this point, static fast waves with wavevectors parallel to ${\boldsymbol
{V}}_p$ in the central object's frame
can exist (i.e., eq. [\ref{waves_fast}] with $\omega =0 \,,
{\boldsymbol {k}} \parallel {\boldsymbol {V}}_p$ is
satisfied). 
The classical fast-magnetosonic point is equivalently defined by the condition
that the poloidal proper speed equals the comoving proper phase speed of a fast-magnetosonic wave
whose wavevector is parallel to ${\boldsymbol {V}}_p$ (see eq. [\ref{theorem}]).}
and all the way to the modified fast-magnetosonic singular point
\citep[see][]{V00}.
The modified (and {\em not} the classical) fast-magnetosonic surface
has the property that it is a singular surface for the steady MHD equations
when one solves simultaneously the Bernoulli and transfield
equations \citep[e.g.,][]{B97}.
Only when the magnetic field geometry is given (and
one solves only the Bernoulli equation but not the transfield
one) does the singular surface correspond to the classical
fast-magnetosonic point.
The modified fast-magnetosonic surface coincides with the limiting characteristic,
the ``event horizon'' for the propagation of fast-magnetosonic waves, since only beyond
this surface all the points inside a fast Mach cone have larger fast
Mach numbers than at the origin of the cone.
At smaller distances, for a part of a given cone, the converse is true: the opening angle
of the fast Mach cone becomes progressively larger as one advances inside that part of the cone;
consequently, a small disturbance in the super-fast regime can affect the entire flow.
The situation is similar to that of light propagating close to a Kerr black hole, where
the ergosphere (which corresponds to the classical fast-magnetosonic
surface) marks the boundary between the elliptic and hyperbolic
regimes, and where the singular event horizon (which corresponds to the modified
fast-magnetosonic surface) is located within the hyperbolic
regime: only for a spherically symmetric (Schwarzschild) black
hole are the ergosphere and event horizon equivalent --- in
direct analogy with spherically symmetric
flows, in which the classical and modified fast-magnetosonic points
coincide \citep[see][]{STT}.

In the case where the outflow attains a highly relativistic speed, the
centrifugal acceleration cannot play an important role. This is
because the nonnegligible $V_{\phi}$ that would be required in
this case would constrain the maximum value of the poloidal
speed: $V_p^2={c^2(1-1/ \gamma^2) - V_{\phi}^2} <{c^2 - V_{\phi}^2}$.
Therefore, in equation (\ref{momentum-V_p}), ${\mbffone}_{C
\parallel} \approx 0$ (and also ${\mbffone}_{G_{\phi} \parallel}\approx 0$).
The ${\mbffone}_{P \parallel}$ force can be neglected since
${\mbffone}_{T \parallel}/{\mbffone}_{P \parallel}=\left(\gamma V_p /c\right)^2 \gg 1$.
The two remaining terms are ${\mbffone}_{T \parallel}$
(a force with a relativistic origin) and ${\mbffone}_{B
\parallel}$. 
The expressions for these terms in equation (\ref{momentum-V_p})
(or, equivalently, eq. [\ref{int-mu}] for the total
energy-to-mass flux ratio) indicate that the bulk Lorentz factor
can increase in response to the decline in either the enthalpy-to-rest-mass ratio
$\xi$ (the thermal acceleration case) or the Poynting-to-mass
flux ratio $\propto -\varpi B_{\phi}$ (the magnetic acceleration
case) along the flow. When the temperature is relativistic, the initial
acceleration is dominated by the temperature force, but after
$\xi$ drops to $\sim 1$ the magnetic force takes over: this is
the likely situation in GRB outflows (see VK01 and \S~\ref{soln_a}).

When the outflow speed is only mildly relativistic, the
magnetocentrifugal force may be important during the initial
acceleration phase, especially if the temperatures
are nonrelativistic; this is the situation envisioned in the
``cold'' relativistic-MHD solutions of \citet*{LCB92}. It is,
however, also conceivable that the magnetic pressure-gradient
force dominates from the start, as might be the case if the
azimuthal field component at the disk surface is large enough;
this is the picture outlined in the presentation of the relativistic
solutions derived by \citeauthor{C94} (\citeyear{C94}; see also
Paper II).

We can obtain an expression for $f_{C \parallel}$ as follows.
By eliminating $\xi \gamma $ between equations (\ref{int-L}) and (\ref{int-mu}),
we obtain a relation between $\varpi V_\phi$ and $\varpi B_\phi$:
\begin{equation}
\frac{ \varpi \Omega V_\phi}{c} = 
1- \frac{\mu(1-x_A^2)}{\mu + \varpi \Omega B_\phi / \Psi_A c^2}\,,
\end{equation}
whose divergence along the flow implies
\begin{equation}
\frac{\partial (\varpi V_\phi)}{\partial \ell} = 
\frac{\mu(1-x_A^2)}{\Psi_A \xi^2 \gamma^2} 
\frac{\partial (\varpi B_\phi)}{\partial \ell} \,.
\end{equation}
Employing the relations for ${\mbffone}_{B \parallel}$, ${\mbffone}_{C \parallel}$
(see eq. [\ref{momentum-V_p}]), and equations (\ref{A1})--(\ref{rho_P}), we obtain
\begin{equation}\label{f_c-f_B}
{\mbffone}_{C \parallel}=
-\frac{\xi \rho_0 \gamma^2 }{2} \frac{\partial V_{\phi}^2}{\partial \ell} +
\frac{(1-x_A^2)(1-M^2-x^2)}{1-M^2-x_A^2} 
\frac{V_{\phi}}{V_p} \frac{B_p}{(-B_{\phi})}
{\mbffone}_{B \parallel}\, .
\end{equation}
The first term on the right-hand side of equation
(\ref{f_c-f_B}) can give rise to either acceleration (when $V_{\phi}$ decreases along the
flowline) or deceleration (when $V_{\phi}$ increases, as in the
corotation regime at the base of the outflow).
This term, together with ${\mbffone}_{G_{\phi} \parallel}$,
can lead to a situation in which $V_p$ increases (resp.,
decreases) and $V_{\phi}$ decreases (resp., increases) while the
Lorentz factor remains roughly constant.
The second term on the right-hand side of equation
(\ref{f_c-f_B}), which is proportional to ${\mbffone}_{B
\parallel}$, demonstrates that the centrifugal force also
has a magnetic component and accounts for the
Poynting-to-kinetic energy transfer that underlies the
magnetocentrifugal acceleration process (see also \citealt{CL94}
for a related discussion).
The form of this term makes it clear why the centrifugal force exceeds the
magnetic force 
during the initial stage of the acceleration, when the flow is still nonrelativistic
($x\ll 1$, $M \ll 1$, with $B_p > |B_{\phi}|$, $V_\phi > V_p$).

The conclusion from the above analysis is that, even though
centrifugal and thermal effects could dominate initially, the
magnetic force eventually takes over and is responsible for the
bulk of the acceleration to high terminal speeds.
\citet{LCB92} described the efficient conversion of
Poynting-to-kinetic energy fluxes in relativistic MHD outflows
in terms of a ``magnetic nozzle'' (see also \citealt{C89}). 
The preceding discussion makes it clear that, in essence,
this mechanism represents the ability of a collimated hydromagnetic outflow
to continue to undergo magnetic acceleration all the way up to
the modified fast-magnetosonic point (which could lie well beyond the
classical fast point). When interpreted in these terms, it
is evident that this effect is not inherently relativistic ---
this conclusion has, in fact, been verified explicitly in the
case of the nonrelativistic self-similar solutions constructed by \citet{V00}.

\subsubsection{Collimation}\label{sectioncollimation}

The projection of equation (\ref{momentum-f}) in the direction
perpendicular to the poloidal flow is given by equation (\ref{forces-perp}).

The two largest terms are the magnetic and electric force components,
which almost cancel each other.

\begin{center}
\plotone{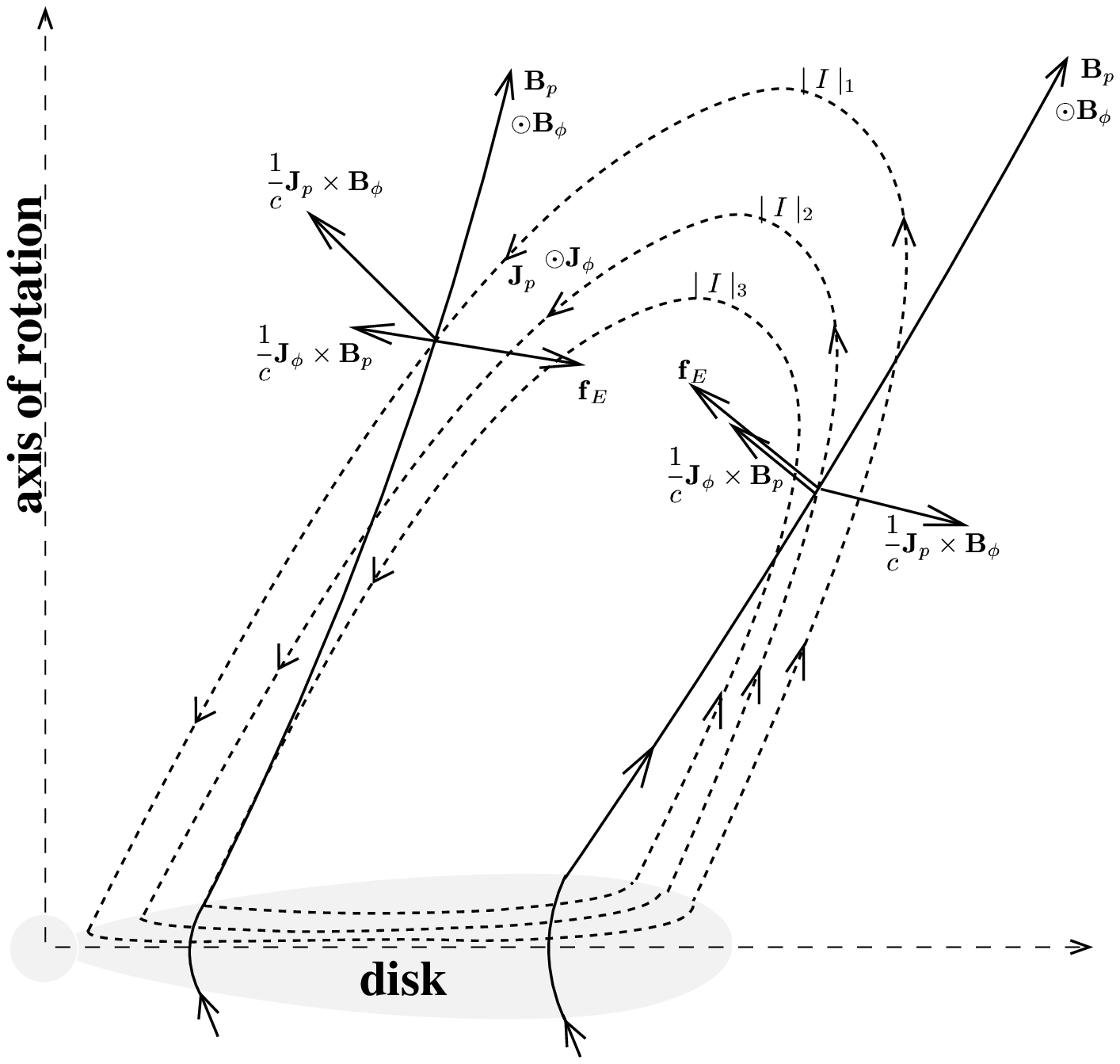}
\figcaption[]
{Sketch of two meridional fieldlines ({\it solid})
and three meridional current lines ({\it dashed}). The currents satisfy
$\mid I \mid_1< \mid I \mid_2 < \mid I \mid_3$.
Given that ${\boldsymbol {J}}_p=\frac{c}{4 \pi } \nablas 
\times {\boldsymbol {B}}_{\phi} = \frac{1}{2 \pi \varpi}
\nablas I \times \hat{\phi}$,
with $I=\int \! \! \! \! \int {\boldsymbol{J}}_p \cdot d {\boldsymbol {S}}
=\frac{c}{2} \varpi B_{\phi}$, the meridional current lines
represent the loci of constant total
poloidal current ($I=const$).
The magnetic and electric forces are shown for the
current-carrying ($J_{\parallel} < 0$, {\it
left}\/ fieldline) and return-current ($ J_{\parallel} > 0$, 
{\it right}\/ fieldline) cases.
\label{fig1}}
\end{center}

The magnetic force in the meridional plane has two parts:
$\frac{1}{c} {\boldsymbol J}_p \times {\boldsymbol B}_{\phi}=
-\frac{B_{\phi}}{4 \pi \varpi} \nablas (\varpi B_{\phi})$ and
$ \frac{1}{c} {\boldsymbol J}_{\phi} \times {\boldsymbol B}_p$.
The first term (which usually dominates) has components
in both the flow (along $\hat{b}\equiv {\boldsymbol B}_p/B_p$)
and the transfield (along $\hat{n}\equiv 
{\boldsymbol E}/E
$) directions. The $\hat{b}$ component 
${\mbffone}_{B \parallel} = -\frac{J_{\bot} B_{\phi}}{c}$
contributes to the acceleration when $J_{\bot}>0$,
where the subscript $\bot$ denotes the vector component along $\hat{n}$.
(If a thermally dominated acceleration regime exists near the base of
the flow, it is in principle possible to have $J_{\bot}<0$
there, corresponding to an enthalpy-to-Poynting energy transfer
mediated by the magnetic force;
see Paper II.) The $\hat{n}$ component $\frac{J_{\parallel} B_{\phi}}{c}$
acts to collimate or decollimate the flow depending on whether
the outflow is, respectively, in the current-carrying ($J_{\parallel}<0$)
or the return-current ($J_{\parallel}>0$)
regime (see Fig. \ref{fig1}).
The second term in the decomposition of the meridional magnetic
force is related to the curvature radius ${\cal R}$,
$\frac{1}{c} {\boldsymbol J}_{\phi} \times {\boldsymbol B}_p=
\hat{n} \frac{B_p^2}{4 \pi} \left(\frac{1}{{\cal R}} -\hat{n} \cdot 
\nablas \ln \mid \nablas A / \varpi \mid \right)$,
and is directed along $\pm \hat{n}$ for $J_{\phi} \lessgtr 0$.

The electric field always points in the $\hat{n}$ direction, but the
electric force ${\mbff}_E = (J^0/c)E\hat{n}$
could lie along either $+ \hat{n}$ or $- \hat{n}$, depending on the sign
of the charge density $J^0/c$. By employing the curvature radius
${\cal R}$, one can write ${\mbff}_{E}=-\hat{n} \frac{x^2 B_p^2}{4 \pi} 
\left(\frac{1}{{\cal R}} -\hat{n} \cdot 
\nablas \ln \mid x \nablas A \mid \right)$.
When ${\cal R} > 0$ (a collimating flow), the effect of the
curvature term in the expression for the electric force is to
oppose collimation, but it is possible for the other term in
this expression to dominate, leading to ${\mbffone}_{E \bot} > 0$.
For highly relativistic motion ($\gamma_p \gg 1$),
$J^0=-\frac{1}{4 \pi} \nablas \left(
{\boldsymbol {V}} \times {\boldsymbol {B}}\right)
=\frac{V_p}{c} J_{\parallel} +\frac{V_{\phi}}{c} J_{\phi}-
\frac{1}{4 \pi} {\boldsymbol {B}}_{\phi} \cdot  \left(
\nablas \times {\boldsymbol {V}}_p\right)
-\frac{1}{4 \pi} {\boldsymbol {B}}_p \cdot  \left( 
\nablas \times {\boldsymbol {V}}_{\phi}\right)
\approx J_{\parallel}$, and the electric force has the sign of
$J_{\parallel}$. In this case ${\mbff}_{E}$ acts to
decollimate the flow in the current-carrying regime
and to collimate it in the return-current regime (see Fig. \ref{fig1}).

For comparison, note that, when the motion is nonrelativistic
($x\ll 1$), ${\mbff}_{E}$ is negligible relative to ${\mbff}_{B}$.
In this case, a flow in the current-carrying regime is
easily collimated (with ${\mbffone}_{B \bot}$ balancing the inertial force
$-{\mbffone}_{I \bot}={\gamma^2 \rho_0 \xi V_p^2}/{{\cal R}}$,
resulting in a nonnegligible value of ${\cal R}$).
In the return-current regime, collimation (resp., decollimation)
is produced if $-J_{\phi} B_p + J_{\parallel} B_{\phi} > 0$
(resp., $<0$); see \cite{O01}.

\section{The $r$ Self-Similar Model}
\label{r-ss_section}

\subsection{Model Construction}
\label{construction}

To obtain semianalytic solutions of the
highly nonlinear system of equations (\ref{bernoulli}) and
(\ref{transfield}), it is necessary to make additional
assumptions: in particular, we look for a way to effect a
separation of variables.

The most complicated expression is the one for $B_{\phi}$ (eq. [\ref{A1}]).
In view of the importance of the azimuthal field component, which
plays a crucial and varied role as part of the magnetic
pressure-gradient, magnetic tension, and centrifugal
acceleration terms in the momentum
equation, the only realistic possibility of deriving exact analytic solutions
is to assume that the $M=const$, $G=const$, and $x=const$
surfaces coincide, i.e., that $M=M(\chi)\,, G=G(\chi) \,, x=x(\chi)$ \citep{V98}.
We aim to find appropriate forms for the functions of $A$
such that the expressions (\ref{bernoulli}) and (\ref{transfield})
become ordinary differential equations (ODEs).
{}From an inspection of the Bernoulli equation (\ref{bernoulli})
we conclude that, in order to separate the variables $\chi$ and $A$
and get a {\em single} equation that only has a $\chi$ dependence,
it is necessary to assume that the $(\nablas A)^2$ term
is a product of a function of $A$ times a function of $\chi$.
As $A$ is a function of $\varpi/G(\chi)$ (see eq. [\ref{G}]),
there must exist functions ${\cal H}_1\,, {\cal H}_2$ such that
$$ \left[\nablas \left( \frac{\varpi }{ G}\right)\right]^2 =
{\cal H}_1 \left(\frac{\varpi }{ G}\right) \ {\cal H}_2 (G)\,.$$
There always exist the trivial possibilities $G\propto r$
in spherical coordinates $(r\,, \theta \,, \phi)$
[$A=A(\theta)$ when the field is radial], and $G=G(\varpi)$
($A=A(\varpi)$ for a cylindrical field), which are not of interest here.
After some algebra one can prove that the only nontrivial
case is to have $G=G(\theta)$, i.e., $\chi=\theta$. It thus
appears that, to obtain an analytic 
adiabatic
solution, it is necessary to
assume $r$ self-similarity.

The remaining assumptions for constructing an $r$ self-similar solution are
that the cylindrical distance (in units of the Alfv\'enic
lever arm), the poloidal Alfv\'enic Mach number, and
the relativistic specific enthalpy are also functions of $\theta$ only:
$x=x(\theta)$, $M=M(\theta)$, $\xi=\xi(\theta)$ (with the result
for $\xi$ following from the nonlinearity of the expression for
$M$; see eq. [\ref{x-M}]).

Following the algorithm described in \citet{VT98}, we change variables
from ($r\,, \theta$) to ($\alpha \,, \theta$) (see
eq. [\ref{alpha}]) and
obtain the forms of the integrals under the assumption of
separability in $\alpha$ and $\theta$.
The results are given in equation (\ref{integrals}).\footnote{The nonrelativistic limit
of our model is {\em not} the 
generalization of the \citet{BP82} model, examined, e.g., in \citet{V00}.
The nonrelativistic limit can, however, be obtained from the analysis of \citet{VT98}:
it corresponds to the third line of their Table 3
(setting $x_1=F-2\,,x_2=F-5/2\,,E_2=C_1=D_2=0$, and ignoring gravity, so it
is possible to assume a polytropic equation of state).}

Among the five unknown functions of $\theta$ 
[$G\left(\theta\right)$, $\psi\left(\theta\right)$,
$M\left(\theta\right)$, $x\left(\theta\right)$,
and $\xi\left(\theta\right)$]\footnote{
Note that these quantities also have an $s$ dependence; see \S~\ref{timerecovery}.}
there are three algebraic equations (eqs. [\ref{x-rss}],
[\ref{M-rss}], and [\ref{bernoulli-rss}]) and two first-order
ODEs (eqs. [\ref{G-rss}] and [\ref{transfield-rss}]).
After solving for these functions, the physical quantities can be recovered using
\begin{mathletters}\label{quantities-rss}
\begin{equation}\label{B-rss}
\frac{\boldsymbol{B}}{B_0 \alpha^{\frac{F-2}{2}}}=
\frac{\sin \theta }
{G^2 \sin \left(\theta - \vartheta \right)} \hat{b}
-\frac{ \mu x_A^4 (1-G^2)}{F \sigma_M x (1-M^2-x^2)} \hat{\phi} \,,
\end{equation}
\begin{equation}\label{E-rss}
\frac{\boldsymbol{E}}{B_0 \alpha^{\frac{F-2}{2}} }=
\frac{ x_A \sin \theta}
{G \sin \left(\theta - \vartheta \right)} \hat{n}
\,,
\end{equation}
\begin{equation}\label{V-rss}
\frac{{\boldsymbol{V}}}{c}=\frac{ F \sigma_M M^2\sin \theta}
{\gamma \xi x^2 \sin \left(\theta - \vartheta \right)} \hat{b}
+\frac{ x_A \mu (G^2-M^2-x^2)}{\gamma \xi G (1-M^2-x^2)} \hat{\phi} \,,
\end{equation}
\begin{equation}\label{gamma-rho-rss}
\gamma=\frac{\mu}{\xi} \frac{1-M^2-x_A^2}{1-M^2-x^2} \,, \quad
\rho_0=\frac{B_0^2 x_A^4 \xi}{4 \pi c^2 F^2 \sigma_M^2 M^2} \alpha^{F-2}\,,
\end{equation}
\begin{equation}\label{P-rss}
P=\frac{B_0^2}{4 \pi} \frac{\Gamma -1}{\Gamma}  \frac{x_A^4}{F^2 \sigma_M^2}
\frac{\xi \left(\xi-1\right)}{M^2} \alpha^{F-2}\,,
\end{equation}
\end{mathletters}
\noindent
where $\hat{b}\equiv {\boldsymbol{B}}_p / B_p =
\hat{z}\cos \vartheta + \hat{\varpi} \sin \vartheta
= \hat{r} \cos \left(\theta - \vartheta \right) -
\hat{\theta} \sin \left(\theta - \vartheta \right)$
is the unit vector along the poloidal fieldline,
$\hat{n} \equiv -{\boldsymbol {\nabla}} A / \mid {\boldsymbol {\nabla}} A \mid =
\hat{z} \sin \vartheta - \hat{\varpi} \cos \vartheta=
-\hat{r} \sin \left(\theta - \vartheta \right) -
\hat{\theta} \cos \left(\theta - \vartheta \right)$ (already
defined in \S \ref{governing}) is the unit vector in the
transfield direction in the meridional plane (toward the axis of rotation),
and $\vartheta \equiv \pi / 2 - \psi$ is the opening half-angle of the outflow
(the angle between the poloidal fieldline and the axis of rotation).

\begin{center}
\plotone{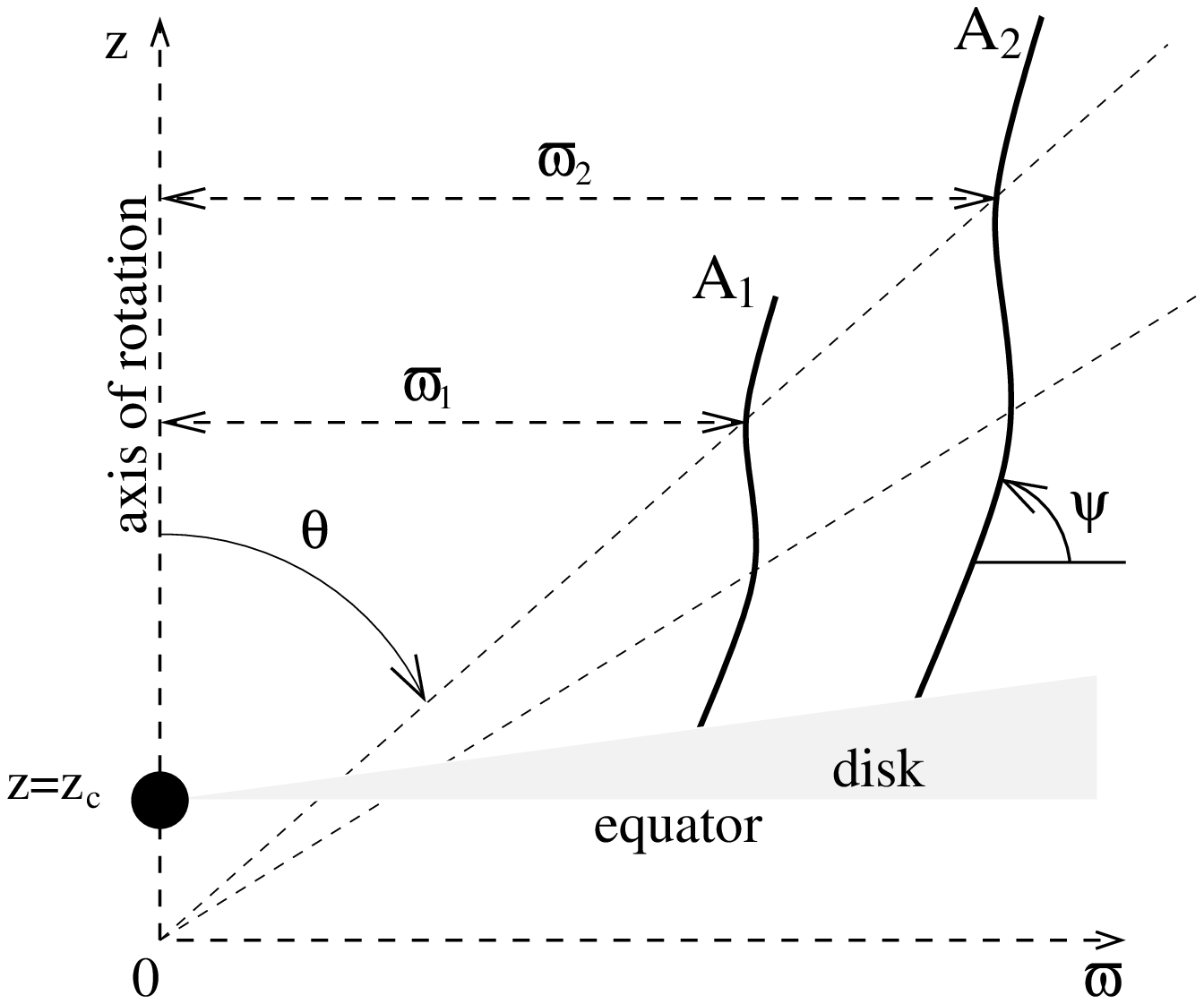}
\figcaption[]
{Sketch of $r$ self-similar fieldlines in the meridional plane.
For any two fieldlines $A_1$ and $A_2$,
the ratio of cylindrical distances for points corresponding to a
given value of $\theta$ is the same for all the cones $\theta=const$:
$\varpi_1/\varpi_2 = (\alpha_1/\alpha_2)^{1/2}$.
\label{fig2}}
\end{center}

The $r$ dependence of all the physical quantities can be
inferred from the expressions (\ref{quantities-rss})
on the basis of the known $r$ dependence of $\alpha$ ($\propto
r^2$; see eq. [\ref{alpha}]). This is a general characteristic
of $r$ self-similar models.

The parameters of the model are $F$ and $\Gamma$ ($=4/3$ in this
study), whereas the ``constants'' 
$x_A(s)$, $\mu(s)$, $\sigma_M(s)$, $q(s)$, and
$B_0(s) \varpi_0^{2-F}(s)$ together with two ``initial'' conditions
(corresponding to the two first-order ODEs) are related to seven boundary
conditions, as we prove in Appendix \ref{appendixrss}.

The $r$ self-similar character of the poloidal field-line shape is shown
in Figure \ref{fig2}.

The model described above is the generalization
to a ``hot'' ($\xi > 1$) gas of the
only known exact semianalytic solution of the relativistic MHD
equations, the ``cold'' $r$ self-similar wind solution found
independently by \citet{LCB92} and \citet{C94}.
(The force-free model of \citealt{C95} can be regarded as a
special case of the latter solutions, corresponding to $M=0$ and
$x_A=1$.)\footnote{The $r$ self-similarity was first employed by \citet{BB78}, who
examined purely HD flows, but it has become well-known only after
\citet{BP82} used it to construct a nonrelativistic MHD
disk-wind model. The latter work has subsequently been generalized by many authors
(see \citealt{V00} and references therein).} 

\subsection{Singular Points}
\label{singular_pts}

\subsubsection{Alfv\'en Singular Point}
\label{alfven-section}
It is obvious from equations (\ref{quantities-rss}) that the
Alfv\'en point, where $G^2=M^2+x^2=1$, is singular.
At this point
\begin{equation}\label{alfven-dispersion}
\left(\gamma V_\theta \right)^2=\frac{B_\theta^2\left(1-x^2\right)}{4 \pi \rho_0 \xi}\,.
\end{equation}
In fact, as ${\boldsymbol{V}}_p \parallel {\boldsymbol{B}}_p$,
this relation holds not only for the 
$\theta$-components of (${\boldsymbol{V}} \,,{\boldsymbol{B}}$),
but for components in any direction in the meridional plane.
Thus, on the Alfv\'en surface, 
static Alfv\'en waves with wavevector in any direction in the meridional plane 
(in the central object's frame) can exist (i.e.,
eq. [\ref{waves_alfven}] with $\omega =0$ and $k_\phi=0$ is satisfied).\footnote{
An equivalent statement is that the Alfv\'en surface marks
the locus of points where the flow proper velocity in any direction in the meridional plane
is equal to the comoving proper phase speed of an Alfv\'en wave that propagates in that direction
(see eq. [\ref{theorem}]).}

In order for the solution to pass through the Alfv\'en singular point,
the Alfv\'en regularity condition
(\ref{alfv2})
must be satisfied.
The latter determines the slope of the ``Alfv\'enic'' Mach number at the Alfv\'en point,
$\left(dM^2/d\theta\right)_A \equiv p_A$,
which is related to the Alfv\'enic value of the magnetization function $\sigma_A$
(see Appendix \ref{appendixrss}).

It is seen from equation (\ref{alfven-dispersion}) that the Alfv\'en point is always located
inside the light surface $x=1$. Note, however, that if $x_A
\approx 1^{-}$ (corresponding to the force-free limit, $M\approx 0$), the Alfv\'en and 
light surfaces almost coincide.

\subsubsection{Magnetosonic Singular Points}
It is straightforward to obtain an expression for $d\psi / d\theta$
as a function of $dM^2 / d \theta$ using the derivative of the
Bernoulli equation (\ref{bernoulli-rss}).
After substituting in the transfield equation (\ref{transfield-rss}), the latter takes the form
$dM^2/d \theta= {\cal N / \cal D}$, where the denominator can be written as
\begin{equation}\label{D}
{\cal D}=
\left(\gamma \frac{V_\theta}{c} \right)^4 - \left(\gamma \frac{V_\theta}{c} \right)^2 
\left(\frac{U_s^2}{c^2}+\frac{B^2-E^2}{4 \pi \rho_0 \xi c^2}\right)
+\frac{U_s^2}{c^2} \frac{B_\theta^2\left(1-x^2\right)}{4 \pi \rho_0 \xi c^2} \,,
\end{equation}
with $B$ being the magnetic field amplitude and with the square
of the proper sound speed given by
\begin{equation}
U_s^2=c^2 \frac{(\Gamma-1)(\xi-1)}{(2-\Gamma)\xi +\Gamma-1}= \frac{ c_s^2 }
{1-{c_s^2}/{c^2} }\, , \quad c_s^2=\Gamma \frac{P}{\rho_0 \xi}\ .
\end{equation}
Singular points appear wherever ${\cal D} =0$;
these are the modified slow and 
fast-magnetosonic 
singular points.
They correspond to points where static 
slow/fast-magnetosonic waves 
with wavevectors along $\hat{\theta}$ in the central object's
frame can exist (i.e., 
eq. [\ref{waves_fast}] with $\omega =0 \,, {\boldsymbol {k}}
\parallel \hat{\theta}$ is satisfied).\footnote{
An equivalent statement is that, at these singular points,
the magnitude of the flow proper velocity along $\hat{\theta}$ is equal
to the comoving proper phase speed of a slow/fast-magnetosonic wave propagating along $\hat{\theta}$
(see eq. [\ref{theorem}]).}
The modified singular surfaces, which correspond to the
``limiting characteristics'' of the self-similar
flow, have previously been considered in connection with the nonrelativistic
solutions (e.g., \citealt{BP82,TSSTC96,B97,V00}).
In order for the solution to pass through a singular point where ${\cal D} =0$,
${\cal N} =0$ must simultaneously hold (yielding the respective regularity condition).

\subsection{Boundary Conditions and Numerical Integration}

When solving the steady, axisymmetric, ideal MHD equations under
the assumption that the azimuthal electric field (as measured in
the central object's frame) vanishes ($E_{\phi}=0$),
seven integrations are required, corresponding to seven unknowns:
the gas density and pressure, the three components of the velocity, and
two functions related to the magnetic field (e.g., $A$ and $B_{\phi}$).

Correspondingly, seven boundary conditions determine a unique solution.
Five of them are the integrals $\Psi_A$, $\Omega$, $L$, $\mu$, and $Q$,
which, as discussed in \S \ref{governing}, are conserved
quantities along the meridional fieldlines.\footnote{
These quantities can be regarded as Riemman invariants; the corresponding
characteristics all coincide with the meridional fieldline.}
The other two correspond to ``initial conditions'' on the
functions $G$ and $M$, which are obtained by integrating
equations (\ref{bernoulli}) and (\ref{transfield}).

In a physically viable solution, the flow starts with a
sub--slow-magnetosonic velocity and must satisfy the causality principle: any disturbance in
the asymptotic regime cannot influence the solution near the
origin through magnetosonic or Alfv\'en waves.
Since the flow starts with a small velocity, it must cross three singular surfaces:
the modified slow-magnetosonic, the Alfv\'en, and the modified
fast-magnetosonic.\footnote{As noted in \S \ref{pol_accel}, the latter surface
represents the ``event horizon'' for the propagation of the fastest waves.
The Alfv\'en surface plays a similar role for the Alfv\'en waves, but
the slow-magnetosonic singular surface is not the ``event horizon'' for the propagation of
slow-magnetosonic waves; it is just the limiting characteristic in the sub-slow hyperbolic
regime (see \citealt{V98}).}
The related three regularity conditions are effectively three boundary conditions
that must be satisfied in order for the solution to pass
smoothly through the singular points. Implementing this
procedure is a highly nonlinear task, since the positions of the singular points are not
known a priori and must be obtained simultaneously with the solution.
All in all, only four boundary conditions remain free and can be specified (e.g.,
on a surface near the origin of the flow).

In the $r$ self-similar model that we examine,
in which we end up integrating ODEs in the variable $\theta$,
it is convenient to choose as the initial surface a cone
$\theta=\theta_i$ (where here and in what follows, a subscript
$i$ denotes an initial value).
One can start the integration by specifying seven initial
conditions (i.e., seven functions of $r$) on this cone (where
$r$ is the arclength along the conical surface; it should not to be
confused with the distance from the central object, which is
given by $[\left(r \sin \theta_i \right)^2 + \left( r \cos
\theta_i - z_c \right)^2]^{1/2}$ ---  see Fig. \ref{fig2}).
For example, one can specify $F\,, \Gamma$ and ${\cal C}_j>0\,, j=1,\dots,7$ such that
$B_{\theta}(r\,,\theta_i)=-{\cal C}_1 r^{F-2}$,
$B_{\phi}(r\,,\theta_i)=-{\cal C}_2 r^{F-2}$,
$V_r(r\,,\theta_i)/c={\cal C}_3 $,
$V_{\theta}(r\,,\theta_i)/c=-{\cal C}_4$,
$V_{\phi}(r\,,\theta_i)/c={\cal C}_5$,
$\rho_0(r\,,\theta_i)={\cal C}_6 r^{2(F-2)}$, and
$P(r\,,\theta_i)={\cal C}_7 r^{2(F-2)}$
(with the ${\cal C}_2,\dots,{\cal C}_7$ being in general functions of $s$).
Note that, in the framework of this self-similar model, the
specified functions of $r$ must be consistent with equations
(\ref{quantities-rss}); if they are not, the given scalings will
not be reproduced on subsequent ($\theta>\theta_i$)
cones.\footnote{
This is a good potential test for numerical codes solving steady-state equations:
starting with the specified forms of the physical quantities on a cone, they
must reproduce the self-similar solution.}
By inverting the system of equations (\ref{quantities-rss}), one can obtain
$x_A$, $\sigma_M$, $q$, $G(\theta_i)$, $M(\theta_i)$, $\psi(\theta_i)$, and $B_0 \varpi_0^{2-F}$
(see eqs. [\ref{invert}]).\footnote{
One can infer $\mu$ from eq. [\ref{bernoulli-rss}] and use it in place of
$\psi(\theta_i)$ as a fieldline constant.}
Three of these parameters are adjusted to satisfy the
regularity conditions at the three singular points. We recall in
this connection that
$\Gamma$ and $F$ are regarded in our formulation as model
parameters (see \S \ref{construction}), and we note that $\psi(\theta_i)$,
$x_A$, $\sigma_M$, and $q$ correspond to the fieldline constants
$\mu$, $x_A$, $\Psi_A$, and $Q$, respectively.

Next we describe our numerical approach.
We have found it most convenient to start the integration from the
Alfv\'en point, since this makes it easier to satisfy the Alfv\'en regularity condition.
We choose a small angular interval $0<d\theta \ll 1$ and specify
the model parameters $F\,, \Gamma$ together with
the following six parameters:
$\theta_A$, $\psi_A$, $x_A$, $\sigma_M$, $q$ and $B_0 \varpi_0^{2-F}$.
(The latter parameter does not appear in the system of equations [\ref{theta-eqs}],
but it affects the magnitudes of the electromagnetic field,
density, and pressure through eqs. [\ref{quantities-rss}].)
The seventh parameter is $p_A$, which is given from the Alfv\'en regularity condition
(see Appendix \ref{appendixrss}).
We start the integration from $\theta=\theta_A \pm d \theta$,
setting $M^2=1-x_A^2 \pm p_A d \theta$,
$G^2=1 \pm {2 \cos \psi_A} {\sin^{-1} \theta_A \cos^{-1}
(\psi_A+\theta_A)} d \theta$, and $\psi = \psi_A$.
Using the upper (lower) signs, we integrate upstream (downstream) from the Alfv\'en point.
Before the first step, we evaluate the parameter
$\mu$ (which is used along the integration path to yield $\psi$)
from equation (\ref{bernoulli-rss}). Whenever the solution hits a singular point,
we adjust one of the above six parameters until a smooth crossing is achieved.

In our model we have ignored gravity, and consequently we expect the flow to be nonsteady
in the sub--slow-magnetosonic regime. We therefore do not attempt to
obtain steady trans--slow-magnetosonic solutions, and thus we do
not continue the integration 
upstream of the slow-magnetosonic
point. This does not, however, affect our ability to
study magnetic effects, as these only become important
downstream of this singular point.

\subsubsection{The Roles of $\Gamma$, $F$, and $z_c$}
\label{roles}

The polytropic index $\Gamma$ controls the thermodynamics of the flow.
For adiabatic flow problems such as the one considered in this paper,
$\Gamma=4/3$ or $5/3$ depending on whether the
temperature is relativistic or not.

The exponent $F$ controls the current distribution.
The poloidal current lines are $I=\frac{c}{2} \varpi B_{\phi}=
const$, where, by equation (\ref{B-rss}),
\begin{equation}\label{current}
I= -c \varpi_0 B_0 \frac{\mu x_A^3}{2 F \sigma_M}
\frac{1-G^2}{1-M^2-x^2} \alpha^{\frac{F-1}{2}} \ .
\end{equation}
Thus, for $F>1$, the current $\mid I \mid$ 
is an increasing function of $\alpha$ near the base
of a trans-Alfv\'enic flow (where $\frac{1-G^2}{1-M^2-x^2} \approx 1$),
corresponding to the current-carrying
regime (see the left
fieldline in Fig. \ref{fig1}). The larger the value of $F-1$, the
stronger the magnetic pinching force
that collimates the flow, and hence the faster the collimation.
The case $F<1$ corresponds to the return-current regime
(represented by the right fieldline in Fig. \ref{fig1}), whereas
$F=1$ corresponds to radial meridional current lines.

In this paper (as well as in Paper II) we concentrate on the case
$F>1$, which should provide a good representation of the
conditions near the axis of highly collimated flows such as GRB
jets. However, in view of the inherent simplifications of the
self-similar formulation, this choice is not unique. For
comparison, we present an $F<1$ solution in \S \ref{soln_d}, and
we also employ a solution of this type in a
forthcoming publication (N. Vlahakis \& A. K\"onigl, in
preparation) in which we model relativistic AGN outflows.
A realistic global solution would encompass both the current-carrying and
return-current regimes, as sketched in Figure \ref{fig1}. This
situation might be mimicked with the help of two, properly
joined, self-similar models: 
one (with $F>1$) that applies near the axis, 
and the other (with $F<1$) that applies further out (at larger cylindrical distances).

The parameter $z_c$
(the $z$ 
coordinate of the disk in the given
system of coordinates; see Fig. 2) affects only the boundary conditions on the disk.
For example, $z_c$ can be used to mimic a Keplerian rotation
law. (Recall from \S \ref{introduction} that, in the
relativistic case considered here,
one cannot naturally incorporate a Keplerian rotation law as in the
nonrelativistic $r$ self-similar solutions.)
In our model, $\Omega = c x(\theta)/\varpi$; thus, for 
$z_c=0$, $\Omega \propto 1/\varpi$ along the conical surface
of the disk. However, for $z_c > 0$, points on the surface of the disk
at different cylindrical distances $\varpi$ correspond to
different values of $\theta(\varpi)$
(see Fig. \ref{fig2}), and $\Omega$ decreases faster than $\propto 1/\varpi$
(with the rate depending on how fast the function $x$ decreases
along the surface of the disk).

\section{Results}\label{results}
The solutions we present in this section are motivated by the
GRB outflow scenario outlined in \S \ref{introduction}.
We approximate the outflow from a disk around a stellar-mass black hole
as a pair of ``pancakes'' \citep[see, e.g.,][]{P99} that move in
opposite directions away from the disk surfaces.
The flow originates from the inner part of the disk, which extends
from the last stable orbit 
around the black hole at $r_{\rm in}$ to
an outer radius $r_{\rm out}$, which for definiteness we choose
to be $3 r_{\rm in}$.
For simplicity we set $z_c=0$, so (see eq. [\ref{alpha}]) $\alpha_{\rm out}/\alpha_{\rm in}=9$.
We take into account the baryonic matter, the electron-positron/photon
fluid, and the large-scale electromagnetic field.

As noted in \S \ref{roles}, we focus on solutions in the
current-carrying regime ($F>1$). For this choice of $F$, equation
(\ref{current}) implies that the current vanishes smoothly
as the axis ($\alpha=0$) is approached. 
A general property of our solutions is that the flow reaches an
asymptotic cylindrical region where the acceleration terminates
(see \S~\ref{soln_a}).
Since we seek to maximize the acceleration efficiency, we consider flows that do
not collimate too fast (see \S~\ref{sectioncollimation}), and
hence we focus on the smallest possible values of $F$.
We therefore choose $F\gtrsim 1$.

Near the origin, the thermal energy associated with the radiation and
$e^\pm$ pairs is nonnegligible; the optical depth is then large enough
to ensure local thermodynamic equilibrium.
We therefore assume that the gas (consisting of
baryons with their neutralizing electrons as well as photons
and pairs) evolves adiabatically.
The polytropic index is fixed at $\Gamma=4/3$, corresponding to relativistic temperatures
for the matter and to a blackbody distribution for the radiation.
Using the Stephan-Boltzmann constant $3 P_{\rm R} / T^4
=3 P/ T^4 ( 1+P_{\rm M}/P_{\rm R})$ (where $T$ is the temperature)
and equations (\ref{P-rss}) and (\ref{M-rss}),
the temperature in units of the electron rest energy is
\begin{equation}\label{Theta}
\Theta = \frac{x_A (\xi-1) 
\alpha^{(F-2)/4}
}{q^{1/4} (1+P_{\rm M}/P_{\rm R})^{1/4} (F \sigma_M)^{1/2}}
\left(\frac{B_0}{1.25 \times 10^{13} {\rm G}}\right)^{1/2} \,,
\end{equation}
or, equivalently,
$\xi=1+(1+P_{\rm M}/P_{\rm R})\Theta^4 \left(\rho_0/
1.39 \times 10^4\ {\rm g \ cm^{-3}}\right)^{-1} \,.$
The matter-to-radiation pressure ratio $P_{\rm M}/P_{\rm R}$ 
is constant in the large-temperature limit $\Theta \gg 1$,
where, given that the pair number density greatly exceeds the baryon number density,
the pair distribution may be approximated by a Maxwellian with zero
chemical potential. In this case $P_{\rm M}/P_{\rm R} = 180/\pi^4 \approx 1.85$ 
(a value very close to the more accurate 
$P_{\rm M}/P_{\rm R} = 7/4 = 1.75$ that characterizes a Fermi distribution).
In the large-temperature limit the pair number density has the
$\Gamma=4/3$ polytropic scaling
($\propto \Theta^3$), but at $\Theta \lesssim 1$ it decreases exponentially,
resulting in $P_{\rm M}/P_{\rm R} \ll 1$. Our polytropic model captures both limits
but not the intermediate temperatures. Which of the two
approximations ($P_{\rm M}/P_{\rm R} =1.85$ or $0$) is more
accurate depends on the initial temperature $\Theta_i$.
For $\Theta_i \gtrsim 1$ we choose $P_{\rm M}/P_{\rm R} =1.85$,
which yields the correct pressure--temperature
relation during the initial stage of the flow,
when thermal effects are important.
This choice introduces an error (leading to an underestimate of the Lorenz factor)
in regions where $\Theta<1$ but
the radiation energy is nonnegligible in
comparison with the baryon rest energy. However, because of the
weak (a power of 1/4) dependence
of $\Theta$ on $1+P_{\rm M}/P_{\rm R}$ in equation
(\ref{Theta}), this error remains small.
We note in this connection that we also neglect the pair
rest-energy density, since it is much smaller than the
matter pressure in the $\Theta\gg 1$ regime where the pair
contribution is maximized.

The mass-loss rate in the outflow is
$\dot{M} = 2 \int \! \! \! \! \int \gamma \rho_0 {\boldsymbol{V}} \cdot d {\boldsymbol {S}}=
\int_{A_{\rm in}}^{A_{\rm out}} \Psi_A \ dA$, or,
\begin{equation}\label{massloss}
\dot{M} =
\frac{B_0^2 x_A^2 \varpi_0^2}{2 F \sigma_M c} \times
\left\{ \begin{array}{lcl}
\displaystyle
\frac{\alpha_{\rm out}^{F-1}-\alpha_{\rm in}^{F-1}}{F-1} &,& F\neq 1 \\
&& \\
\displaystyle
\ln \frac{\alpha_{\rm out}}{\alpha_{\rm in}} &,& F = 1\, .
\end{array}\right. 
\end{equation}
If $\Delta t$ is the burst duration, then $M_b\approx \dot{M} \, \Delta t$ is the
total baryon mass ejected.
The total energy is ${\cal E}_i \approx \mu M_bc^2$, and initially
it resides predominantly in the electromagnetic field; the initial thermal
energy is approximately $\xi_i M_b c^2 \approx
(\xi_i / \mu) {\cal E}_i$.\footnote{Different shells have different 
baryon mass densities, so the more accurate expressions are
$M_b= \int \dot{M}(s) \ ds /c$, ${\cal E}_i = \int \mu(s) \dot{M}(s) c^2 \ ds /c$,
and (for the initial enthalpy-to-kinetic-energy ratio)
$\int \xi_i(s) \gamma_i(s) \dot{M}(s) c^2 \ ds /c$.}

In VK01 we considered two lower limits on the baryon loading,
corresponding, respectively, to the requirements
(1) that the flow be optically thick in the region where the pair number density
drops below that of the baryons and (2) that the flow be
matter-dominated when it becomes optically thin.
We also obtained an upper limit on $\rho_0$
from the requirement that the flow be optically thin in the
internal GRB-shock regime.
In the solutions presented in this paper we evaluate the 
optical depths more accurately. Specifically, consider two
neighboring fieldlines labeled by $A$ and $A+dA$.
Along a direction $\hat{\zeta}$ that makes an angle $\omega_{\zeta}$ to
the flow velocity, the optical depth is \citep{ANP91}
\begin{equation}
d \tau=\gamma \left(1-\frac{V}{c} \cos \omega_{\zeta} \right)
n_e \sigma_T \frac{d \zeta_{\bot}}{\sin \omega_{\zeta}}\,,
\end{equation}
where $n_e$ is the electron/positron number density, $\sigma_T$
is the Thomson
cross section, and $d \zeta_{\bot}$ is the distance between the
fieldlines. The optical depth is minimized when $\omega_{\zeta} = \arcsin (1/\gamma)$,
for which $d \tau = n_e \sigma_T d \zeta_{\bot}$
(corresponding to photons moving perpendicular to the flow direction in the
comoving frame).
Starting from a point on the inner fieldline, we chart the
photon trajectory by using
$\omega_{\zeta} = \arcsin (1/\gamma)$ and $d \tau = n_e \sigma_T d \zeta_{\bot}$
until the outer fieldline is reached, and use this information
to evaluate the optical depth.

\begin{figure*}
\centering
  {\includegraphics[width=1.\textwidth]{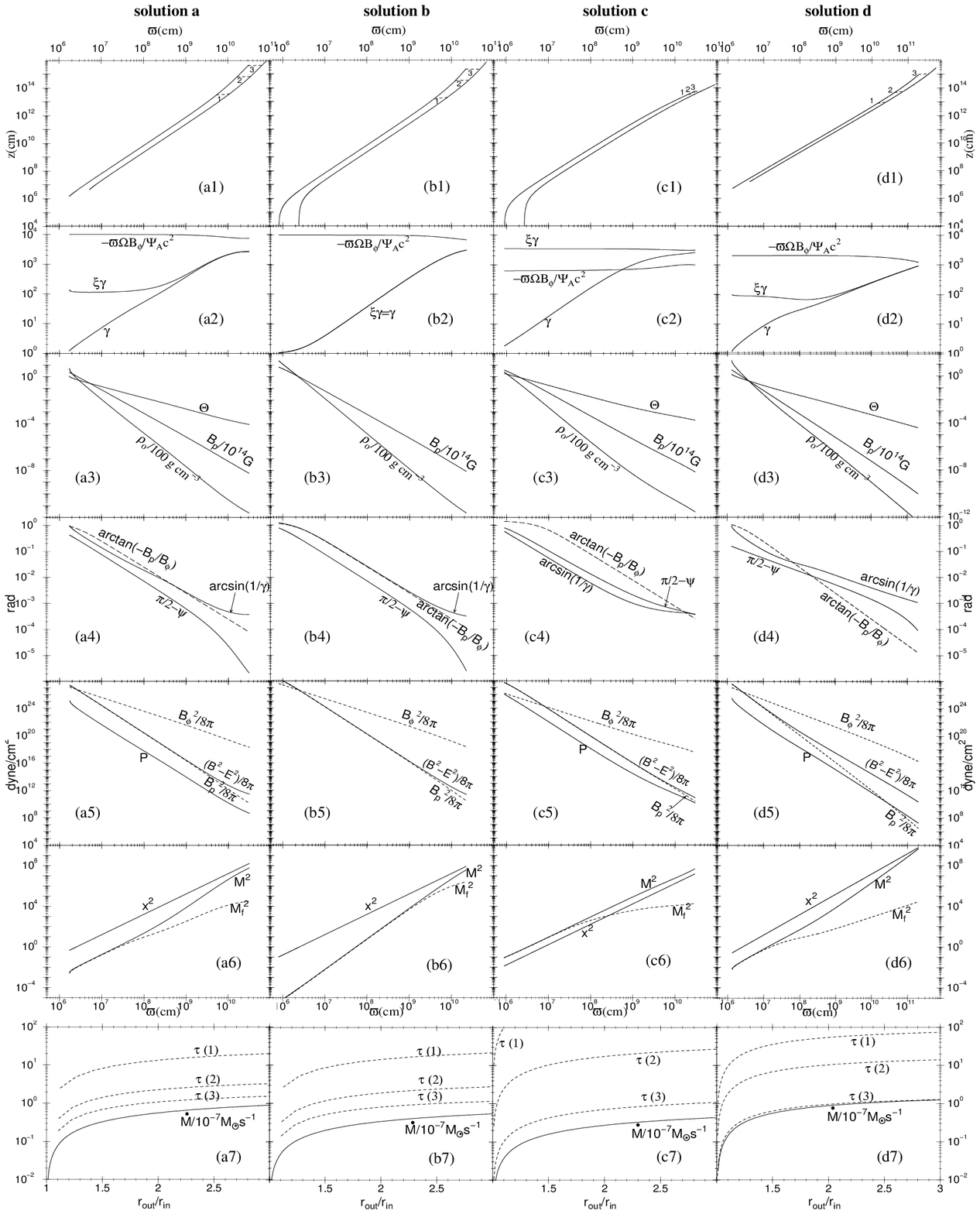}}
  \caption{
Main properties of the four representative solutions discussed
in \S \ref{results}. See text for details.
\label{fig3}}
\end{figure*}

We now present the results of the numerical integration for four
representative solutions (labeled $a$, $b$, $c$, and
$d$), for which the parameters are given in Table \ref{table1}.
The most important physical quantities are plotted in Figure \ref{fig3},
in which each column corresponds to a given solution.
The properties of these solutions are described in
detail in the following subsections.

\begin{table*}
\begin{center}
\caption{
Parameters of Representative Solutions$^\dagger$
\label{table1}}
{
\scriptsize
\begin{tabular}{lccccccccccc}
\tableline\tableline
solution & $F$ & $x_A^2$ & $\sigma_M$ & $q$ & $\mu$ & $B_0 \varpi_0^{2-F}$(cgs) &
$\theta_A(\degr)$ & $\psi_A(\degr)$ & 
$p_A$ &
$\sigma_A$
& $\varpi_0$($10^6$cm)
\\
\tableline
\quad
$a^*$
 & $1.01$ & $0.99$ & $5000$ & $37.396$ & $10116.1$ & $1.35 \times 10^{20}$ &
$35$ & $72.7$ & 
$-3.9 \times 10^{-2}$ &
$86.6$
& $2.5$ \\
\quad
$b$ & $1.01$ & $0.9999$ & $5000$ & $0$ & $9997.4$ & $1.05 \times 10^{20}$ &
$35$ & $72.7$ & 
$-5.0 \times 10^{-4}$ &
$6820$
& $2.5$ \\
\quad
$c$ & $1.01$ & $0.15$ & $300$ & $322501$ & $4053.2$ & $5.93 \times 10^{19}$ &
$35$ & $72.0$ & 
$-3.0 $ &
$0.18$
& $3$ \\
\quad
$d$ & $0.7$ & $0.96$ & $1000$ & $44.92$ & $2150$ & $6.83 \times 10^{21}$ &
$10$ & $83.4$ & 
$-0.93 $ &
$23$
& $2.5$ \\
\tableline
\multicolumn{11}{l}{
$^\dagger$ In all cases $\Gamma = 4/3\,, z_c=0$, and $r_{\rm out}/r_{\rm in}=3$.} \\
\multicolumn{11}{l}{
$^*$
The solution presented in VK01
is the same as solution $a$, except that
$B_0 \varpi_0^{2-F}=2.96 \times 10^{19}$ cgs and $\varpi_0=7.8125 \times 10^5$ cm.
}\\
\end{tabular}
}
\end{center}
\end{table*}

\subsection{Solution $a$: A Hot, Fast-Rotator Outflow}
\label{soln_a}
This solution represents our fiducial model of a
trans-Alfv\'enic GRB outflow in which the Poynting flux exceeds
the enthalpy flux at the origin ($\mu \gg \xi_i$).
The flow starts from the disk with a nonrelativistic velocity and
in a short distance crosses the slow magnetosonic point (where $V_p \approx c/\sqrt{3}$).
The acceleration in this regime is due to the pressure gradient force
(the centrifugal acceleration is negligible for $\xi_i \gg 1$).
The slow magnetosonic point arises from the interplay between
the vertical gravitational and pressure-gradient forces. As we neglect gravity, we start
the integration slightly above that slow point
($\gamma \gtrsim (3/2)^{1/2}$, see Fig. \ref{fig3}a2).

In this initial sub-Alfv\'enic regime, $M_i^2 \ll 1-x_A^2\,, G_i^2 \ll 1\,, x_i^2\ll x_A^2$.
Equation (\ref{A2}) for the Lorentz factor gives
\begin{equation}\label{ximu}
\xi_i \gamma_i (1-x_i^2) \approx \mu (1-x_A^2) \,,
\quad \mbox{or} \quad \xi_i \approx \mu (1-x_A^2)\,.
\end{equation}
Thus, a Poynting-dominated flow ($\mu \gg \xi_i$) is always
close to being force-free ($x_A \approx 1$), and the initial
enthalpy-to-Poynting flux ratio satisfies
$\xi_i/\mu \approx 1-x_A^2$ ($=0.01$ in the displayed solution).

As the flow moves downstream it crosses the Alfv\'en singular
point. We solve numerically for the slope  $p_A$ of the Alfv\'enic Mach number that satisfies
the Alfv\'en regularity condition (see \S \ref{alfven-section}).

In the super-Alfv\'enic regime there are three possible cases:\\
1. The flow recollimates ($\psi > \pi/2$)
and there is a termination point where the entire energy is suddenly
transformed into kinetic motion (corresponding to the solution hitting the modified
fast singular surface but not being able to cross it); \\
2. The flow starts to decelerate at some heigh above the disk; \\
3. The flow becomes asymptotically cylindrical.

The last case is the only physically acceptable one and corresponds to a
specific value of one of the model parameters. (For example, for
the chosen parameter set $F$, $\theta_A$, $p_A$,
$x_A$, $\sigma_M$, and $q$, there is a unique value of $\mu$ that corresponds
to the cylindrical solution).
This is the ``magnetic nozzle'' mechanism first described by
\citet{LCB92} and discussed in \S \ref{pol_accel}.
When the flow has cylindrical asymptotics, the asymptotic regime $\theta = 0$
is the only possible solution of ${\cal D}=0$ (where ${\cal D}$ is given by
eq. [\ref{D}]). In this case, the modified fast-magnetosonic singular
surface (the ``event horizon'' for the propagation of fast waves)
corresponds to the asymptotic cylindrical regime (where all the fast Mach cones along
each fieldline have the same opening angle).
As in the ``cold'' solutions derived by \citet{LCB92}, the critical value of $\mu$ is always
close to $2 \, \sigma_M$.

{}Figure \ref{fig3}a1 shows the meridional projections
of the innermost and outermost fieldlines on a
logarithmic scale. Our chosen initial cylindrical distances are
$\varpi_i \approx 1.7 \times 10^6$ cm
for the innermost fieldline and $\approx 5.2 \times 10^6$ cm
for the outermost one; these distances are only slightly larger
than those of the footpoints of these fieldlines in the disk.
Figure \ref{fig3}a1 also depicts the optical paths of photons
that originate at three points on the innermost fieldline;
these were calculated according to the procedure described
at the beginning of this section and are plotted as
dashed lines labeled 1, 2, and 3. Figure \ref{fig3}a7 shows the
corresponding optical depths $\tau(1)$, $\tau(2)$, and $\tau(3)$
along these paths as a function of $r_{\rm out}/r_{\rm in}=
(\alpha_{\rm out}/\alpha_{\rm in})^{1/2}$. It is seen that, as
one moves downstream (with the initial point on the innermost
fieldline shifting to a larger height $z$ above the 
disk), the optical depth gets
smaller, and the flow eventually becomes optically thin
($\tau(3) \approx 1$ for $r_{\rm out}/r_{\rm in}=3$).

\begin{center}
\epsscale{0.96}
\plotone{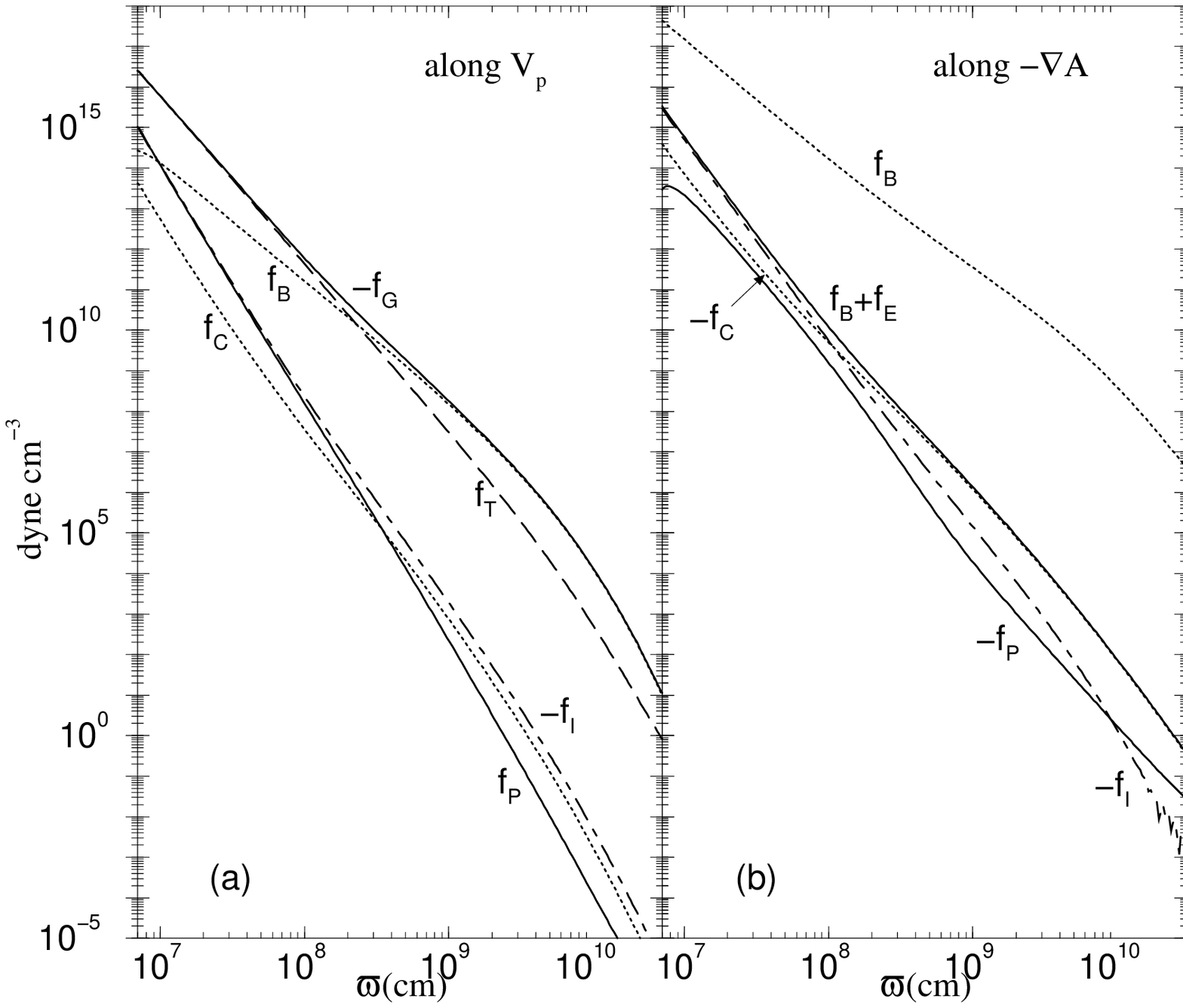}
\figcaption[]
{Force densities in the meridional plane (a) along the poloidal flow
and (b) in the transfield direction (toward the axis)
for solution $a$.
\label{fig4}}
\end{center}

{}Figure \ref{fig3}a7 also shows the radial profile of the mass-loss rate.
For $r_{\rm out}/r_{\rm in}=3$,
$\dot{M} \approx 10^{-7}\ M_{\sun}\ {\rm s}^{-1}$,
corresponding to a total ejected baryonic mass of $M_b \approx 10^{-6}\, M_{\sun}$
for a typical burst duration $\Delta t \approx 10\, {\rm s}$.

The total energy flux along the poloidal flow is 
$\mu c^2 \times \gamma \rho_0 V_p$ (where $\gamma \rho_0 V_p$
is the mass flux).
The main part of the total energy flux (at least in the initial phase of the flow)
is the Poynting flux (see eq. [\ref{tensor_flux}]),
\begin{equation}
c \left(T^{0j}_{\rm EM} \hat{x}_j \right) \cdot \frac{{\boldsymbol {V}}_p}{V_p}=
\frac{c}{4 \pi} \left({\boldsymbol {E}} \times {\boldsymbol {B}}\right) \cdot \frac{{\boldsymbol {V}}_p}{V_p} = 
-\frac{\varpi \Omega B_{\phi}}{\Psi_A } \times \gamma \rho_0 V_p\,.
\end{equation}
(Note that this term is positive since $B_{\phi} < 0$.)
The remaining part $(\mu c^2 + \varpi \Omega B_{\phi} / \Psi_A ) \times \gamma \rho_0 V_p = 
\xi \gamma c^2 \times \gamma \rho_0 V_p$ (see the
energy conservation relation, eq. [\ref{int-mu}])
includes the kinetic energy flux ($\gamma \rho_0 c^2 (\gamma -1)
V_p$) and the enthalpy flux.
The total energy loss rate is
$\dot{{\cal E}_i}=2 \int \! \! \! \! \int \gamma \rho_0 V_p
\left(\xi \gamma c^2 -\varpi \Omega B_{\phi} / \Psi_A \right) 
\frac{{\boldsymbol{V}}_p}{V_p} \cdot d {\boldsymbol {S}}= \mu \dot{M} c^2$,
and for $\Delta t \approx 10$s
the total energy injected into the two oppositely directed jets is
${\cal E}_i \approx 1.8\times 10^{52}\, {\rm ergs}$.

{}Figure \ref{fig3}a2 shows the various energy fluxes in units of
$\gamma \rho_0 c^2 V_p$ (the mass flux $\times c^2$)
as well as the Lorentz factor $\gamma$
as functions of $\varpi$, the distance from the axis of rotation
along the innermost fieldline.
We distinguish three different regimes:

1. \underline{Thermal acceleration region:}\\
{}From the initial point (slightly above the slow point) up to $\sim 10^8$ cm,
$\xi \gamma \approx \xi_i = const.$ 
(The exact initial values for solution $a$ are
$\gamma_i = 1.2$ and $\xi_i = 114.8$.)
The acceleration is primarily thermal: enthalpy is transformed into kinetic energy.
In Figure \ref{fig4}a we see that the force $-{\mbffone}_{G \parallel}$ (which
measures the increase in $\gamma$) is equal to the temperature
force ${\mbffone}_{T \parallel}$
(which describes the decrease of $\xi$).
The Poynting-to-mass flux ratio ($- \varpi \Omega B_{\phi} / \Psi_A $) is 
essentially constant, which means that the field is force-free; it only
guides the flow.
To a good approximation, $\gamma \approx \gamma_i \varpi /
\varpi_i$ and $\xi \approx \xi_i \varpi_i / \varpi $.
So long as $\xi \gg 1$, equation (\ref{Theta}) gives $\xi \propto \Theta$;
thus, $\Theta \approx \Theta_i \varpi_i / \varpi$.
This is verified in Figure \ref{fig3}a3, which shows the poloidal magnetic
field (in units of $10^{14}$ G) and the baryon rest-mass density (in units of $100$ g cm$^{-3}$).
It is seen that 
$\rho_0 \propto 1/\varpi^3$ (as expected from the polytropic relation $P \propto \rho_0^{4/3}$) and
$B_p \propto 1/\varpi^2$ (as expected from the constancy of the mass-to-magnetic flux ratio
$4 \pi \gamma \rho_0 V_p/B_p = \Psi_A$).
As the azimuthal velocity is negligible (for a highly relativistic poloidal motion),
equation (\ref{int-Omega}) implies $\varpi B_{\phi} = const.$,
as expected in the force-free regime.
In general, $V_{\phi}=cx+V_p B_{\phi}/B_p$, so $-B_{\phi} \approx xB_p =E$,
consistent with $B_p \propto 1/\varpi^2$ and $B_{\phi} \propto 1/\varpi$.
\\
{}Figure \ref{fig3}a4 shows the angle between the total magnetic field and its
azimuthal component (which, given that $B_p \propto 1/\varpi^2$ and $-B_{\phi} \propto 1/\varpi$,
decreases as $1/\varpi$),
the ``causal connection'' opening angle $\arcsin (1/\gamma) \propto 1/\varpi$,
and the opening half-angle of the outflow $\vartheta = \pi/2 - \psi$
(its initial value is $\vartheta_i \approx 25\degr$).
\\
{}Figure \ref{fig3}a5 shows the pressures associated with the
poloidal, azimuthal, and comoving magnetic field components,
as well as the total thermal (lepton + radiation) pressure contribution
($B_p^2 / 8 \pi \,, B_{\phi}^2 / 8 \pi$ --- dashed lines;
$B_{\rm co}^2 /8 \pi = (B^2-E^2)/ 8 \pi \,, P$ --- solid lines, respectively). 
In general, $B_{\rm co}^2 \approx B_p^2 + B_{\phi}^2 / \gamma^2
\approx B_p^2 \left(1+ x^2 /  \gamma^2\right)$.
As $x$ and $\gamma$ are both proportional to $\varpi$, their ratio is a
constant, $x/ \gamma = x_i / \gamma_i \approx x_i \ll 1$.
Thus, $B_{\rm co} \approx B_p$, as verified by the figure.
\\
{}Figure \ref{fig3}a6 shows $x^2 = \left(\varpi \Omega / c
\right)^2$ and $M^2 =  \left(\gamma V_p \right)^2 / \left(B_p^2
/ 4  \pi \rho_0 \xi \right)$ (the squares of the ``light
cylinder'' radius and the Alfv\'enic Mach number, respectively)
as well as the square of the ``classical fast-magnetosonic proper Mach number''
$M_f^2 \equiv \left(\gamma V_p \right)^2 / U_f^2 $,
where $U_f$ is the larger solution of the quadratic
\begin{equation}\label{cl_fast}
\left(\frac{U_f}{c} \right)^4 - \left(\frac{U_f}{c} \right)^2
\left(\frac{U_s^2}{c^2}+\frac{{B}^2-{E}^2}{4 \pi \rho_0 \xi c^2}\right)
+\frac{U_s^2}{c^2} \frac{B_p^2\left(1-x^2\right)}{4 \pi \rho_0 \xi c^2}=0\,.
\end{equation}
The point where $x=1$ corresponds to the light surface, which is close to
the Alfv\'en surface $x=x_A$ (more accurately, the Alfv\'en surface
is where $M^2=1-x_A^2$).\\
The point $M_f=1$ is the classical fast-magnetosonic 
point. It is seen that, up to that point, $M \approx M_f$. This can be
understood by noting that, since $B^2-E^2 \approx B_p^2 \gg 4 \pi \rho_0 \xi U_s^2$,
the solution of equation (\ref{cl_fast}) yields
$M^2 / M_f^2= 1/2+(1/4 + x^2 M^2 U_s^2 / \gamma^2 V_p^2)^{1/2}$.
It follows that, so long as $x^2 M^2 U_s^2 / \gamma^2 V_p^2 
\le x_i^2 M^2 /2 \ll 1/2$ (using $U_s^2 \le c^2/2$), $M_f \approx M$.\\
At the classical fast point (subscript $f$) $M\approx 1 \ll x$, so equation (\ref{A2})
for the Lorentz factor gives
$\gamma_f \approx \mu / \xi_f x_f^2$.
If this point is located inside the thermally dominated region
($\xi_f \gg 1$, as in the depicted solution), then, using $\gamma_f \xi_f \approx
\xi_i \gamma_i$, one infers 
$x_f \approx (\mu / \xi_i \gamma_i)^{1/2}$.
As $\gamma / x \approx \gamma_i / x_i$, it follows that
$\gamma_f \approx (\mu \gamma_i / \xi_i x_i^2)^{1/2}$.

2. \underline{Magnetic acceleration region:}\\
{}From the end of the thermal acceleration zone, where $\xi \approx 1$, up to
$\varpi \approx 10^{10}$ cm, it is seen from Figure \ref{fig3}a2
that the Lorentz factor continues to increase as $\gamma \propto \varpi^{\beta}$,
with $\beta$ a constant $\approx 1$.
This exponent is in general different for different solutions.
We find that for solutions with a given $\mu$ but different $\xi_i$,
$\beta $ is always less than 1 and decreases with increasing $\xi_i$.
For larger $\xi_i$ the magnetic effects are less important,
resulting in a weaker collimation of the flow and hence in a
larger asymptotic width $\varpi_{\infty}$.
Since the final Lorentz factor is $\sim \mu/2$, the exponent 
$\beta = d \ln \gamma / d \ln \varpi $ should be smaller.
Centrifugal acceleration can also influence the magnitude of this exponent:
for mildly relativistic flows we expect this acceleration to
increase the lever arm of the flow, resulting in a lower value of $\beta $.
\\
The acceleration in this region is due to magnetic effects:
Poynting energy is transformed into kinetic energy.
{}Figure \ref{fig4}a shows that the force $-{\mbffone}_{G \parallel}$ (which describes
the increase in $\gamma$) is equal to the Lorentz force ${\mbffone}_{B \parallel}$
(which derives from the decrease of $\mid \varpi B_{\phi}\mid $).
The Poynting-to-mass flux ratio declines from its initial value $\approx \mu c^2$ as 
$\mu c^2 - \mid \varpi \Omega B_{\phi}  / { \Psi_A } \mid  \propto \varpi^{\beta}$,
with $\beta \approx 1$, a result verified by Figure
\ref{fig3}a2. However, the deviation from the initial value of $\mu c^2$ is not
too strong, so the scalings $B_p \propto 1/\varpi^2$, $\rho_0 \propto 1/\varpi^3$,
and $-B_p / B_{\phi} \propto 1/\varpi$
remain approximately the same as at smaller radii (see
Figs. \ref{fig3}a3, \ref{fig3}a4, and \ref{fig3}a5).\\
{}Figure \ref{fig3}a6 shows that $x\gg M$ but that $M$ increases faster than $x$.
As the Poynting-to-matter energy flux ratio is $ 
\left[ c \left({\boldsymbol {E}} \times {\boldsymbol {B}}\right)
\cdot {\hat{b}} / 4 \pi \right] / \left[\xi \gamma^2 \rho_0 c^2 V_p\right]=
(\mu - \xi \gamma) / \xi \gamma  = (x^2 - x_A^2) / (M^2 + 1 - x_A^2)
\approx  x^2 / M^2  $,
this is another manifestation of the Poynting-to-kinetic flux conversion.\\
The Bernoulli equation simplifies in this region
to $\gamma \approx \gamma V_p /c$,
which can be used, along with $\gamma \approx \mu M^2 / \xi (x^2+M^2)$ and equation (\ref{V-rss}),
to obtain the fieldline slope,
\begin{equation}\label{bernoulli_asympt}
1-\frac{F \sigma_M}{\mu}\frac{x^2+M^2}{x^2} \approx 
1- \frac{\sin(\theta-\vartheta)}{\sin \theta}
\approx \frac{\vartheta}{\theta}
=\left(\frac{d \ln \varpi}{d \ln z}\right)_{A=const}\ .
\end{equation}
(The same result can be found using eq. [\ref{bernoulli-rss}].)
So long as $x\gg M$, the slope is $1-F \sigma_M / \mu \approx 1/2$ 
(as the critical value for $\mu$ is close to $2 \sigma_M$, and $F \approx 1$).
Thus, the shape of the fieldlines is parabolic, $z \propto \varpi^2$,
as Figure \ref{fig3}a1 verifies \citep[cf.][]{C95}.

3. \underline{Asymptotic cylindrical region}\\
At the end of the magnetic acceleration region the
flow becomes cylindrical: Figure \ref{fig3}a1 shows that the
fieldlines tend to a constant $\varpi$, whereas Figure
\ref{fig3}a4 indicates that the opening half-angle
$\vartheta=\pi/2 - \psi$ tends to zero.\\
{}Figure \ref{fig4}b shows that, although the electric force almost cancels
the magnetic force (${\mbffone}_{B\bot}+{\mbffone}_{E\bot} \ll {\mbffone}_{B\bot}$),
their sum ${\mbffone}_{B\bot}+{\mbffone}_{E\bot}
\approx -{\mbffone}_{C\bot}$.
On the other hand, $-{\mbffone}_{I\bot} \ll -{\mbffone}_{C\bot}$,
or (using eqs. [\ref{fcbot}] and [\ref{fibot}]),
$\varpi / {\cal R} \ll V_\phi ^2 \sin \psi / V_p^2$, which means
that the poloidal curvature radius vanishes (i.e., the poloidal
fieldlines become straight) --- a characteristic of cylindrical collimation.
(In the initial acceleration region near the disk it is seen from
Fig. \ref{fig4}b that ${\mbffone}_{B\bot}+{\mbffone}_{E\bot} 
\approx -{\mbffone}_{I\bot}$, which implies
that the fieldlines curve toward the axis.)\footnote{
The acceleration could in principle continue
if a transition from positive to negative poloidal curvature were possible.
Although the geometry of the $r$ self-similarity does not allow such a transition,
other types of self-similar solutions can be constructed in which such a transition takes
place and the flow continues to accelerate \cite[see][]{V03}.}
\\
In the cylindrical region, the transfield force-balance equation becomes
\begin{equation}
\rho_0 \xi \gamma^2 \frac{V_{\phi}^2}{\varpi}=
\frac{d}{d \varpi} \left(P+\frac{B^2}{8 \pi}\right) +\frac{B_{\phi}^2}{4  \pi \varpi}-
\frac{J^0}{c} E \,, 
\end{equation}
or, using $\alpha \propto \varpi^2$ and equations (\ref{quantities-rss}),
\begin{equation}\label{transf_cyl}
\rho_0 \xi \gamma^2 \frac{V_{\phi}^2}{\varpi} +
2 \frac{2-F}{\varpi} \left(P+\frac{B_p^2}{8 \pi}\right)=
\frac{F-1}{4 \pi \varpi} \left(B_{\phi}^2 - E^2\right)\,.
\end{equation}
As the left-hand side of equation (\ref{transf_cyl}) is small but positive,
$(F-1)  \left(B_{\phi}^2 - E^2\right) / B_{\phi}^2 \approx 0^+$.
Using $B_{\rm co}^2 \approx B_p^2 + B_{\phi}^2 / \gamma^2 $
and $B_{\rm co}^2 = B_p^2 + B_\phi ^2 - E^2$,
we conclude that $B_{\phi}^2$ is invariably $\gtrsim E^2$; hence, only a
current-carrying jet ($F>1$) can have cylindrical asymptotics.
\\
Equation (\ref{bernoulli_asympt}) with $d \varpi=0$ gives 
the final kinetic-to-Poynting flux ratio,
$(M^2/x^2)_{\infty} \approx (\mu/F\sigma_M) -1 \approx 1$,
corresponding to $\gamma_{\infty} \approx \mu - F \sigma_M \approx \mu/2$.
The value of $M^2$ becomes as large as that of $x^2$ (see
Fig. \ref{fig3}a6), so a rough equipartition holds between the
kinetic and Poynting fluxes: $\gamma_{\infty} \approx \left(-\varpi \Omega B_{\phi} / \Psi_A c^2
\right)_{\infty} $ (see Fig. \ref{fig3}a2).
\\
The implied conversion efficiency of 
$\sim 50\%$ between the total 
energy (injected largely in the form of a Poynting flux)
and the final kinetic energy of the flow is consistent with the internal
shock scenario for GRBs. Specifically, the asymptotic kinetic
energy in each jet is $\sim 2\times 10^{51}\, {\rm ergs}$ for
our fiducial parameters. With a radiative efficiency of $\gtrsim
10\%$ \citep[e.g.,][]{KPS97,NP02b}, this implies a radiated $\gamma$-ray
energy of a few times $10^{50}$ ergs, as inferred
observationally \citep{F01}.
\\
{}Figure \ref{fig3}a1 shows that, as the cylindrical region is approached, 
the Lorentz factor increases more slowly than $\propto \varpi$,
resulting in a nonnegligible value of $x / \gamma$.
Since $B_{\rm co}^2 \approx B_p^2 \left(1+ x^2 /  \gamma^2\right)$,
this explains the divergence of the $(B^2-E^2)/8 \pi$ and
$B_p^2 /8 \pi$ curves in Fig. \ref{fig3}a5 near the cylindrical region.
(Eq. [\ref{transf_cyl}] can be written as
$ (F-1) (B^2-E^2) = B_p^2 + 4 \pi \rho_0 \xi \gamma^2 V_{\phi}^2
+ 8 \pi (2-F) P$, which shows that $B^2-E^2 \gtrsim B_p^2 /
(F-1)$ in the cylindrical regime.)
\\
The condition that the GRB emission region be optically thin
to the prompt $\gamma$-ray photons typically implies a lower
limit $\gtrsim 100$ on the Lorentz factor of the outflow (e.g., \citealt{LS01}).
The asymptotic Lorentz factor attained in our fiducial solution
(and, in fact, in the other three solutions presented in this
section) readily satisfies this requirement.

\subsubsection{Time-Dependent Effects}\label{timerecovery}
As noted at the beginning of \S \ref{results},
we approximate the outflow 
as a pair of pancakes that move in opposite directions away from
the disk.
In the context of the internal-shock scenario,
each pancake consists of a number $N=100 \ N_2$ of shells.
The width of each shell is $\delta s = \Delta \ell / N = c \Delta t /N=
3 \times 10^9 (\Delta t)_1 / N_2 $ cm, where the total duration of the burst is
$\Delta t = 10 \ (\Delta t)_1$ s.

If the ejection of the shells from the disk surface ($\ell =0$)
starts at time $t=0$ and ends at time $\Delta t$, then 
each shell can be labeled by its ejection time $
t_i=
s/c$, with
$0\le s \le c \Delta t$.
The shell ``$s$'' moves along the poloidal fieldline
and at time $t$ its position is 
$\ell =\int_{s/c}^t V_p \ dt$.
The distance between two neighboring shells $s$, $s+\delta s$ is
(using $V_p \approx c - c/2 \gamma^2$)
\begin{equation}\label{distance_of_shells}
\delta \ell = \delta \left( \int_{s/c}^t V_p \ dt \right) 
\approx -\delta s + 
\displaystyle \int_0^t  \frac{c}{\gamma^2} \frac{\delta \gamma}{\gamma} dt \,.
\end{equation}
At early times the second term on the right-hand side of
equation (\ref{distance_of_shells}) is negligible
and the frozen-pulse approximation holds (the distance between the two shells
is constant, $\delta \ell \approx - \delta s$).
However, even if the integrand in the second term is tiny,
there will eventually come a time (call it $t_c$) when the
integral will cancel the $-\delta s$
term (for $\delta \gamma / \delta s > 0$, corresponding to the trailing
shell moving faster than the leading one).
One can integrate the equation of motion for each shell, $dz/dt=c
(1-1/\gamma^2)^{1/2}$, with
$\gamma \approx \gamma_i \varpi / \varpi_i  \approx \gamma_i
(z/z_i)^{1/2}$, to show that two neighboring shells with
$\delta \gamma \sim \gamma$
will not collide for as long as they move in the flow
acceleration zone. 
This is a reflection of the fact that the difference in the initial
Lorentz factors of the two shells is not large enough to
compensate for the longer acceleration time of the leading shell
as it crosses this region, which also implies that the frozen-pulse 
assumption continues to hold throughout the acceleration zone.
Only after the shells reach the constant-velocity,
cylindrical-flow region, will the two shells collide
($\delta \ell =0 $). 
Using equation (\ref{distance_of_shells}),
this will happen at a distance $\approx \gamma_{\infty}^2 \delta s 
\approx 2\times 10^{16} (\Delta t)_1 / N_2\ {\rm cm}$
beyond the end of the acceleration zone.

So long as the frozen-pulse approximation is valid, 
$\ell = \int_{s/c}^t V_p \ dt \approx ct - s$.
As we proved in \S \ref{governing}, it is possible to examine the motion of
each shell in this regime using the $r$ self-similar model.
If we focus on the shell $s_0$, then the parameter $F$ and boundary conditions 
$\{x_A(s_0) \, ,\, \mu(s_0) \, ,\, \sigma_M(s_0)\, ,\, q(s_0)\,
,\, B_0(s_0)\varpi_0(s_0)^{2-F}\}$ that we used to specify the solution refer
to this particular $s_0$, and the corresponding solution 
$\{M(\theta , s_0)\,,\, x(\theta , s_0)\}$ is valid only for this shell.

In general, we may choose a different set $\{F \, ,\, x_A(s) \, ,\,
\mu(s) \, ,\, \sigma_M(s) \, ,\, q(s)\, ,\, B_0(s) \varpi_0(s)^{2-F}\}$ for a
different shell $s$, but we have to be careful to also satisfy
the assumption of a quasi-steady poloidal field, which requires
all the shells to experience the same poloidal magnetic field at any given location.
This requirement constrains the choice of boundary conditions.

No constraint is necessary in the force-free regime, in which the
electromagnetic
field is effectively decoupled from the matter
in that there is no feedback from the matter acceleration to the field.
The electromagnetic
field only guides the flow and the motion is effectively HD.
As we proved in VK01, one simply has to replace the 
spherical $r$ in the radial-outflow scalings (e.g., \citealp{P99})
with the cylindrical radius $\varpi$ to get the correct scalings in the
magnetically guided case.

However, in the superfast regime ($M^2 \gg 1-x_A^2$, $x^2 \gg
x_A^2$), one finds that the (appropriately simplified)
Bernoulli and transfield force-balance equations
(eqs. [\ref{bernoulli-rss}] and [\ref{transfield-rss}])
become completely $s$-independent if one writes
$M^2(\theta\,, s)={\cal M}^2(\theta) g(s)$,
$x^2(\theta\,, s)={\cal X}^2(\theta) g(s)$, and
$\mu(s) / \sigma_M(s) = \mu(s_0) / \sigma_M(s_0)$.
($F$ must also be $s$-independent.)
Thus, we are free to specify the functions
$g(s)$, $x_A(s)$, $\sigma_M(s)$, and $q(s)$ when
$\mu(s)=\mu(s_0) \sigma_M(s)/ \sigma_M(s_0)$, and
from the fact that $A$ is $s$-independent we then also have
$B_0(s)\varpi_0^{2-F}(s) x_A^F(s) / g^{F/2}(s) = 
B_0(s_0)\varpi_0^{2-F}(s_0) x_A^F(s_0) / g^{F/2}(s_0)$.
These functions correspond to the initial conditions for each shell,
obeying the quasi-steady poloidal magnetic field assumption.

Using $s=ct-\ell$ in any quantity $\Phi=\Phi(A\,, \ell\,, s)$,
we may find either the time dependence following 
the motion of a particular shell:
$\Phi= \Phi(A\,, ct-s \,, s)$ with $s$ held constant, 
or the time dependence at a given point in space 
as different shells pass by:
$\Phi= \Phi(A\,, \ell \,, ct - \ell)$, with $\ell$ held fixed.
The ``initial'' value $\Phi_i=\Phi(A\,, 0 \,, s) = \Phi(A\,, 0 \,, ct_i)$ 
corresponds to the
time $t=t_i=s/c$ when the shell $s$ is ejected from the surface $\ell=0$.
Thus, we see that the $s$-dependence in $\Phi(A\,, \ell\,, s)$ represents the 
initial conditions at the ejection surface for each shell.

\begin{center}
\plotone{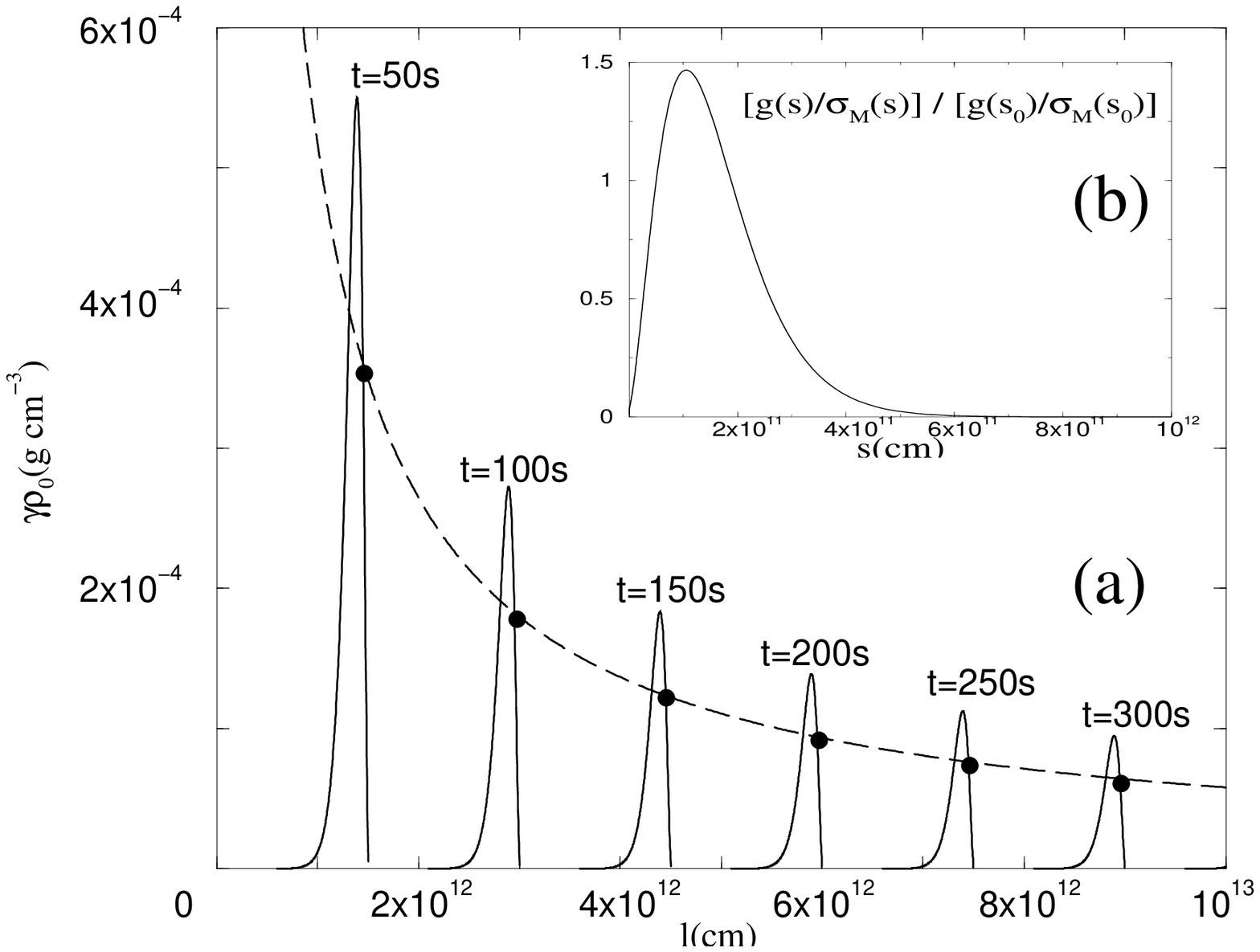}
\figcaption[]
{($a$) Baryon mass density plotted as a function of the arclength along the
innermost fieldline of solution $a$ for several values of time
since the start of the burst.
The pancake width $(\Delta \ell)$ remains constant $\approx c \Delta t$.
The {\it dashed}\/ line represents the ``time independent''
solution for the reference shell
$s_0=5 \times 10^{10}$ cm, corresponding to the {\it heavy}\/ dots.
($b$) Assumed form of the function $g(s)/\sigma_M(s)$ across the pancake.
This function represents the initial baryon mass density in each shell
($\gamma \rho_{0i} (s) \propto g(s)/\sigma_M(s)$; see eq. [\ref{gamma-rho-rss1}]).
The shell $s=0$, which is ejected first, has a tiny mass, the following
shells have progressively larger masses, and the last shells ($s
\lesssim c \Delta t$) again have negligible masses.
\label{fig5}}
\end{center}

To illustrate how the time dependence in our model can be recovered,
we now consider the baryon mass density along the flow.
{}From equation (\ref{gamma-rho-rss}),
\begin{eqnarray}
&& \gamma \rho_0 (r\,, \theta \,, s) =
\frac{\left[B_0(s_0)\varpi_0^{2-F}(s_0)\right]^{\frac{4}{F}}
\mu(s_0) x_A^4(s_0) g(s)
\
A^{2-\frac{4}{F}} 
}{4 \pi c^2 F^{\frac{4}{F}} \sigma_M(s_0) g^2(s_0)
\left[{\cal M}^2(\theta) + {\cal X}^2(\theta)\right] \sigma_M(s)} 
\nonumber \\ && \times 
\frac{{\cal M}^2(\theta) g(s)-1+x_A^2(s)}
{{\cal M}^2(\theta) g(s)}
\frac{{\cal X}^2(\theta) g(s) + {\cal M}^2(\theta) g(s)}
{{\cal X}^2(\theta) g(s) + {\cal M}^2(\theta) g(s)-1}\ .
\end{eqnarray}
Neglecting the second line (which is approximately equal to 1 in
the non--force-free regime),
\begin{equation}\label{gamma-rho-rss1}
\gamma \rho_0 (r\,, \theta \,, s) = \gamma \rho_0 (r\,, \theta \,, s_0)
\frac{g(s)/\sigma_M(s)}{g(s_0)/\sigma_M(s_0)} \,.
\end{equation}
{}Figure \ref{fig5} shows $\gamma \rho_0$ as a function of
the arclength along the inner fieldline at various times
(as in the HD case illustrated in Fig. 1a of \citealt{PSN93}).
At each instant of time $t$, one can identify (using $s=ct - \ell$) which shell $s$
is at the point $\ell$. Knowing the distribution of $\gamma
\rho_0$ as a function of $\ell$ for the reference
shell $s_0$, one can obtain the density for all the
other shells (corresponding to a particular choice of the
function $g(s)/\sigma_M(s)$, such as the one depicted in Fig. \ref{fig5}b).

\subsubsection{Validity of Assumptions}\label{validity}

The ideal-MHD theory is not always a good approximation;
in some cases it must be replaced by an exact multi-fluid theory
including a relativistic Ohm's law \citep[e.g.,][]{MM96}.
A necessary condition for its applicability, derived from the
need to attain the requisite charge density and current,
implies a lower limit on the matter density \citep*[e.g.,][]{SDD01}.
The limits are $J \ll \rho_0 q_e c / m_p$, $J^0/c \ll \rho_0 q_e
/ m_p $ (where $q_e$ the electron charge),
and we have verified that they are well satisfied in all of our solutions.
In particular, the protons and the neutralizing electrons have a sufficiently large
number density to screen the electric field parallel to the flow
and to provide the necessary charge density and current.

Our neglect of gravity has turned out to be a valid
approximation, since the flow near the
disk is thermally dominated and the temperature force density near the origin
of our fiducial solution
(at $\varpi_i \approx 2 \times 10^6\ {\rm cm}$, slightly above the slow-magnetosonic point)
is ${\mbffone}_{T \parallel} \approx 10^{19}$ dyne
cm$^{-3}$, which is much larger than the gravitational force
density associated with a central object of a few solar masses.

In \S~\ref{timerecovery} we demonstrated that the frozen-pulse
assumption is valid throughout the acceleration region.

We have also neglected the terms 
$\partial(E+B_{\phi})/\partial s$ and $\partial(E^2-B_{\phi}^2)/\partial s$ 
in equation (\ref{momentum-s}).
Our results by and large verify these approximations, as we find that
only near the asymptotic cylindrical region does $B_{\rm co} =
(B_p^2 + B_\phi ^2 - E^2)^{1/2}$ deviate from $B_p$ (see Fig. \ref{fig3}a5).
Even so, the ratio of the terms in equation (\ref{momentum-s}) that contain
$E+B_\phi$ over the last term in that equation is
$\gamma (E+B_\phi) / B_\phi$, which (using eq. [\ref{ohm}]) becomes
$\gamma (1-V_p/c) \ll 1$.
An alternative argument supporting the conclusion that 
$E+B_\phi\approx 0$ is based on
the requirements that $B_{\rm co}^2>0$ and $V_\phi>0$.
The former gives (using $B_{\rm co}^2=B_p^2+B_\phi^2-E^2$ and eq. [\ref{E_A}])
$B_\phi^2/E^2 > 1-1/x^2$, whereas the latter yields
(using eq. [\ref{V-B}]) $B_\phi^2/E^2 < c^2/V_p^2$.
Thus, the ratio $B_\phi^2/E^2$ is bounded in a tiny interval around $1$,
from $1-1/x^2$ to $c^2/V_p^2 \approx 1+1/\gamma^2$.

\subsection{Other Solutions}

\subsubsection{Solution $b$: The Cold Case ($\xi_i\approx 1$)}
\label{soln_b}
In the limit of low temperatures ($q \rightarrow 0$, $\xi
\rightarrow 1$), we
reproduce the exact cold relativistic solutions of \citet{LCB92},
using a different parameter regime appropriate to GRB outflows.
\citet{C94} employed the same model but examined more massive
flows that were not close to being force-free and thus were
not accelerated as efficiently.

We present a cold solution in the second column of Figure \ref{fig3};
it can be readily compared with the fiducial solution $a$.

The cold solution is closer to being force-free, since
$\gamma_i \approx 1/ (1-x_i^2)^{1/2}$ (using 
$V_{pi}\approx 0 \,, V_{\phi i}\approx \varpi_i \Omega$), and hence
(by eq. [\ref{ximu}]) $\mu (1-x_A^2) \approx (1-x_i^2)^{1/2}\approx 1$.
(For comparison, $\mu (1-x_A^2) \approx \xi_i$ for solution $a$.)

Since the flow does not pass through a slow magnetosonic point
in this case, we are able to find the
solution all the way down to the disk surface. Near the origin
the flow corotates with the disk ($V_{\phi} \approx cx$) and is accelerated
due to the centrifugal force ${\mbffone}_{C \parallel}$.
The Lorentz factor does not increase linearly with $\varpi$
(but rather, to a good approximation, as $\gamma \approx
1/(1-x^2)^{1/2} \approx (1+x^2)^{1/2}$).
However, the Lorentz force soon takes over and thereafter
$\gamma \propto \varpi$ (see Fig. \ref{fig3}b2).

At the classical fast point
$M_f \approx M \approx 1$, 
equation (\ref{A2}) gives
$\gamma_f \approx \mu / x_f^2$, which (using $\gamma_f \approx x_f$)
can be transformed into the familiar form $\gamma_f \approx
\mu^{1/3}$ \citep[e.g.,][]{C86}.
Unlike the purely monopole
case \citep{M69}, this point is located
at a finite distance from
the origin, and the bulk of the magnetic acceleration occurs 
further downstream.

The flow near the cylindrical regime is similar to that in the
fiducial solution, as thermal effects are not important anymore.
However, a signature of the cold solution can be seen in
{}Figure \ref{fig3}b6 in the weaker deviation of $M$ from $M_f$.
(The solution of eq. [\ref{cl_fast}] in this case is $M_f^2/M^2 = B_p^2 /
(B^2-E^2)$, and the deviation of $M$ from $M_f$ is due only to
$B^2-E^2$ becoming larger than $B_p^2$, which happens near the
cylindrical region.)

\subsubsection{Solution $c$: The Slow-Rotator Case ($\xi_i \approx \mu$)}

This solution describes a slow rotator, in which the Poynting flux is smaller
than the enthalpy flux (so $\mu \approx \xi_i$).
The angular velocity of the inner fieldline is $\Omega = c x_A/\varpi_0 = 
3873$ rad s$^{-1}$ (see Table \ref{table1}),
much smaller than the typical value ($\sim 10^4$ rad s$^{-1}$) 
near a solar-mass black hole.
This solution is characterized by
significantly higher $B_p / \mid B_{\phi} \mid$ ratios
(see Fig. \ref{fig3}c4) and much lower values of $x$ (see Fig. \ref{fig3}c6)
in comparison with solutions $a$ and $b$.
It is also seen that this solution is not force-free: $x_A^2
=0.15 \ll 1$ (see eq. [\ref{ximu}]).
The acceleration in this case is thermal ($\xi \gamma \approx
\mu $; see Fig. \ref{fig3}c2).

Nevertheless, since the flow is trans-Alfv\'enic, the poloidal magnetic
pressure is much larger than the thermal pressure (see Fig. \ref{fig3}c5).
This can be understood from the fact that, in general, one has
at the Alfv\'en point
\begin{equation}
\left(\frac{B_p^2}{ 8 \pi P} \right)_A= \left(
\frac{2}{1-1/\xi} \frac{\left(\gamma V_p/c\right)^2}{1-x^2}\right)_A \approx
\frac{2 \gamma_A^2}{1-x_A^2} \gg 1\,.
\end{equation}
Only when the flow is super-Alfv\'enic everywhere (even near the origin)
can one obtain a hot, radial, HD-like solution with $P > B_p^2 /
8 \pi$.

The magnetic field only guides the flow in this case.
The collimation, however, is not very efficient (see Fig. \ref{fig3}c1).
The flowlines ``attempt'' to become conical, but since in the framework of
$r$ self-similarity the radial distance is given by
$r= \varpi_0 \alpha^{1/2}(A) G(\theta)/ \sin \theta$,
the only possible asymptotes as $r \rightarrow \infty$ are
$\theta \rightarrow \theta_\infty$. For $\theta_\infty=0$
we have cylindrical asymptotics (as in solutions $a$ and $b$),
and the only other possibility is for all the flowlines
to have the same asymptotic opening half-angle, which is the situation
depicted in Figure \ref{fig3}c1. 
This behavior can be understood if we write the transfield force-balance equation as
\begin{eqnarray}\label{transfield_V03}
\frac{B_p^2}{4 \pi {\cal R}}\left(M^2+x^2-1\right)=
-\frac{1}{8 \pi \varpi^2} \hat{n} \cdot {{\nablas}} \left[\varpi^2 \left(B_{\phi}^2 -E^2
\right) \right]
+ \nonumber \\ 
+\frac{B_p^2}{4 \pi \varpi} \left(\frac{M V_{\phi} }{V_p}\right)^2 \hat{n} \cdot \hat{\varpi}
- \hat{n} \cdot \nablas \left(P+\frac{B_p^2}{8 \pi}\right)
\end{eqnarray}
(see \citealp{V03}).
By neglecting the centrifugal ($\propto V_\phi^2$) and the poloidal magnetic pressure terms
and using equation (\ref{ohm}) (which implies $E \approx -B_\phi V_p/c$),
equation (\ref{transfield_V03}) can be rewritten as
\begin{eqnarray}\label{transfield_V03_appr}
\frac{B_p^2}{4 \pi {\cal R}}\left(M^2+x^2-1\right)=
-\frac{1}{8 \pi \varpi^2} \hat{n} \cdot \nablas \left(\frac{\varpi B_{\phi}}{\gamma}
\right)^2 - \hat{n} \cdot \nablas P \,.   
\end{eqnarray}
The first term on the right-hand side is positive and acts to collimate the flow (resulting in
a positive poloidal curvature, ${\cal R} > 0$),
whereas the second term is negative and has the opposite effect on ${\cal R}$.
In Poynting flux-dominated solutions, the pressure term is negligible and the collimation
continues (albeit at a very low rate; see paper II) until the shape becomes cylindrical.
However, when the thermal pressure becomes comparable to the comoving magnetic pressure
(as in the present solution; see Fig. \ref{fig3}c5), the two terms
on the right-hand side of equation (\ref{transfield_V03_appr}) cancel each other.
The curvature $1/{\cal R}$ thus vanishes before the cylindrical regime is reached,
resulting in an asymptotic conical flow.

\subsubsection{Solution $d$: The Return-Current Regime}
\label{soln_d}
For completeness, we also present a return-current ($ J_{\parallel} > 0$) solution.
Since, near the disk surface, $-\varpi
B_{\phi} \propto \alpha^{F-1}$ 
is a decreasing function of $\alpha$ in the $F<1$ case under consideration,
the magnetic force acts to {\em de}collimate the flow
(see \S~\ref{sectioncollimation}).
This results in a weaker collimation than in the current-carrying ($F>1$)
solutions (see Fig. \ref{fig3}d1).
The electric force succeeds in collimating the flow despite the
counter effect of the magnetic force; as discussed in \S
\ref{sectioncollimation}, this is
associated with the positive charge density ($J^0>0$) expected
in the return-current regime. 
Examining the forces in the transfield direction, we find that 
${\mbffone}_{E \bot} \gg {\mbffone}_{E
\bot}+{\mbffone}_{B \bot}
\approx -{\mbffone}_{I \bot} \gg -{\mbffone}_{C
\bot}$; i.e., the Lorentz force is balanced by the inertial
force. (This can be contrasted with the cylindrical solution $a$, where
the balance is provided by the centrifugal force:
${\mbffone}_{B \bot} \gg {\mbffone}_{E
\bot}+{\mbffone}_{B \bot} \approx -{\mbffone}_{C \bot}
\gg -{\mbffone}_{I \bot} $; see Figure \ref{fig4}b). As in
the nonrelativistic, cold MHD flow considered by \citet{BP82},
the curvature radius is nonnegligible and the solution
terminates at a finite height above the disk.

\section{Discussion and Conclusion}
\label{conclusions}
We regard the fiducial solution $a$ as providing the most plausible 
representation of an accelerating
GRB outflow. Neutrino energy deposition and magnetic field
dissipation are likely to give rise to a nonnegligible thermal
component at the origin, as envisioned in the original fireball
scenario. Such a component is incorporated into solution $a$ but
is missing in solution $b$. It is, nevertheless, worth
reemphasizing that the dominant energy source even in solution
$a$ is the Poynting flux. We note in this connection that one of
the potential problems of traditional
fireball models has been the ``baryon contamination'' issue: Why
does the highly super-Eddington luminosity that drives the
fireball not lead to a more significant mass loading of the
flow and thereby keep $\gamma_\infty$ comparatively low? If most
of the requisite energy were injected in a nonradiative form, then
this problem would be alleviated.\footnote{As we argue in
\citet*{VPK03}, the baryon loading
issue may in principle be completely resolved in the context of
the MHD acceleration model.} The expectation that GRB outflows
are Poynting flux-dominated also renders the slow-rotator
(enthalpy flux-dominated) solution $c$ less relevant to their
interpretation than the fiducial solution.

GRB outflows are inferred to be highly collimated at large
distances from the origin: this has motivated us to construct our fiducial
solution in the current-carrying ($J_\parallel
< 0$) regime, which is applicable near the symmetry axis (see
Fig. \ref{fig1}). However, as demonstrated in Figure \ref{fig3},
the illustrative current-carrying solution $d$ also becomes well
collimated on a similar length scale, so it may also
provide an adequate representation of such outflows. In fact, this regime
may be a natural choice for modeling the far-downstream region of the flow
since it corresponds to a plausible current-configuration
on large scales; furthermore, as shown in \citet{V03}, it also results in a more
efficient acceleration.
We have opted to focus on the current-carrying regime in this paper since the
solutions in this case, unlike the return-current solutions 
formally extends to infinity.
We incorporate the return-current regime into a GRB outflow model in \citet{VPK03}.

Our fiducial self-similar ``hot'' solution as well as our
``cold'' relativistic MHD solution
(and the corresponding ones derived by \citealt{LCB92} and \citealt{C94})
have cylindrical asymptotics. In the nonrelativistic theory, it
was shown \citep{VT98} that it is possible to derive exact solutions of radially
self-similar flows that have either one of the following three
types of asymptotes: cylindrical, parabolic, or
conical. Although we expect that it should be possible
to obtain different asymptotic behaviors also in radially
self-similar relativistic flows, this has not yet been studied in detail.
Our basic conclusions about the magnetic acceleration of the
flow do not, however, depend on the shape of the asymptotic
flowlines, although the quantitative constraints on the mass
loading (to assure that the flow is optically thin in the
shell-collision region) would be eased if the flow
continued to expand rather than collimated to a cylinder.
GRB jets are often modeled in terms of conical jets, and
this picture has gained support from observations of a panchromatic break
in the light curve of several GRB afterglows (from which
jet half-opening angles in the range of $2^{\circ}$ to
$20^{\circ}$ have been inferred; e.g., \citealt{PK02}). 
However, a sharp physical boundary may not be required for
interpreting the data in the context of a beam-patterned outflow
\citep[e.g.,][]{RLR02,LB02}. Furthermore, it
has been argued that a cylindrical jet model could provide a better fit
to the light curves of at least some afterglows and might
perhaps even be able to account for the panchromatic breaks commonly
attributed to a finite jet opening angle \citep[e.g.,][]{D98,CHL01}.

Even if GRB jets are not characterized by a global
cylindrical geometry, the asymptotically cylindrical solutions
that we derived may be applicable to the ``patchy shell''
scenario for variable GRB outflows \citep{KP00}. In this
picture, the outflowing shells are ejected within a given
opening angle but do not fill the entire solid angle into which
they are injected. In the context of the internal shock model,
this scenario alleviates the energy
requirements on bright bursts; it is also consistent with the
apparent lack of a strong correlation between the prompt
high-energy and the afterglow fluences, and it may account for the
early-time variability in the afterglow emission of a
source like GRB 021004 \citep{NPG03}. In the phenomenological
model considered by \citet{KP00}, successive blobs of angular
scale $1/\gamma_\infty$ are ejected
randomly within an opening angle of $10^{\circ}$, and the number
of blobs ejected along a given direction (given that more than
one is required for the
production of internal shocks) is also taken to be a random
number (distributed uniformly in a predetermined range). This
model was intended to mimic the behavior of causally
disconnected regions in large-scale ejected shells. An
alternative physical picture is that of a collection of
magnetically active regions on the surface of the disk that feed
the GRB outflow. In this picture, the magnetic field associated
with each ``patch'' guides successive ``mini'' shells along roughly
congruent paths, thereby enhancing the efficiency of
internal-shock production. However, the ejection directions from different
sites need not be parallel to each other, so some of the beamed
$\gamma$-ray emission may miss the observer: this might
contribute to the appearance of quiescent times between GRB pulses
\citep{NP02a}.

The high ($\sim 50\%$) Poynting-to-kinetic energy conversion
efficiency exhibited by all our 
fast-rotator solutions (solutions $a$, $b$, and $d$)
has made it possible for our models to accommodate the
internal-shock emission scenario without requiring
a prohibitively large energy input at the source (see \S
\ref{results}). However, currently available data on
the radiated $\gamma$-ray energy $E_\gamma$ \citep[e.g.,][]{PK02} and the
outflow kinetic energy $E_K$ \citep[e.g.,][]{F01}, which indicate that
they are both approximately equal ($\sim 10^{51}\, {\rm ergs}$),
are inconsistent with this scenario for any reasonable radiative
energy efficiency of such shocks (see \citealt{NP02b} for further
discussion of this issue). One possible
resolution of this apparent inconsistency is that the prompt
high-energy emission arises directly from electromagnetic energy
dissipation without the intermediate step of
Poynting-to-bulk-kinetic energy conversion \citep[e.g.,][]{T94,SU00,LB01,DS02}. 
Our solutions seem to be consistent with this possibility: they
imply that about half of the injected energy is converted into bulk
kinetic energy; if the remaining Poynting flux can be
efficiently converted into high-energy emission, then this would
lead to an approximate equality between $E_\gamma$ and
$E_K$. \citet{DS02} used ideal-MHD equations with parameterized magnetic energy
dissipation to demonstrate (in the context of a strictly radial flow,
i.e., without taking into account the transfield force-balance equation
that determines the fieldline shape)
that the Poynting-to-radiation energy conversion could in
principle be as high as 50\% and that the dissipation may also
result in significant bulk acceleration. It would be interesting
to combine the approach taken by these authors (see also
\citealt{SDD01} and \citealt{D02}) with the one utilized in this
paper to consider the effects of dissipation in a {\em nonradial} flow.

We now briefly compare the results of our exact solutions with
some of the previous work on MHD
effects in GRB outflows. The incorporation of ``disposable''
magnetic energy into the traditional fireball model was
discussed by \citet{MLR93}. 
The behavior that they infer is consistent with our solutions
in the regime where $\gamma \propto \varpi$ (but not beyond
it). \citet{U94} interpreted the outflows
in terms of pulsar winds that generate intense electromagnetic
waves at the point where the ideal-MHD approximation breaks
down. He highlighted the fact that the Lorentz factors of the
particles accelerated by such waves could be as high as $\sim
\mu^{2/3}$ and thus greatly exceed the terminal Lorentz factors 
($\sim \mu^{1/3}$) attained in the \citet{M69} ideal-MHD 
monopole solution. 
The acceleration efficiency of our collimating (and thus nonmonopolar)
ideal-MHD solutions is much higher yet:
they yield $\gamma_\infty \approx \mu/2 \gg \mu^{2/3}$.

\citet{MR97} examined a magnetized conical jet with a pure
electron-positron composition. They deduced (and our solutions have confirmed) that the
pair--photon fluid is initially accelerated by thermal pressure, with
the magnetic field acting only to guide the flow. They also
pointed out that radiation drag effects allow the pairs
and photons to remain coupled even beyond the point where the
scattering optical depth drops below 1 and that at some point pair
annihilation freezes out. \cite{GW98}
subsequently showed that the surviving pairs 
carry only a fraction $\sim 10^{-5} \gamma_i^{3/4}$ of the initial
energy, implying that, in the absence of magnetic acceleration,
this scenario is extremely inefficient.
We note, however, that if a nonnegligible fraction ($1- \gamma_i
\xi_i /\mu$) of the total injected energy is in magnetic form
([with the ratio of magnetic to pairs-plus-photons energy
injection rates being $(\mu -  \gamma_i \xi_i) / \gamma_i \xi_i
\approx \gamma_i$, as assumed by \citealt{MR97}),
then, based on our ``cold'' solution (see \S~\ref{soln_b}), half of the magnetic energy
could be eventually transformed into kinetic energy of the surviving pairs,
reaching overall efficiencies $\sim (\mu - \xi_i \gamma_i)/ 2 \mu$.\footnote{In this 
estimate we identify $\rho_0$ as the rest-mass density of the final pair population, which
represents only a tiny fraction of the initial rest-mass density of pairs.}
For $\gamma_i \sim 1$ this yields an efficiency of $\sim 25 \%$,
which is significantly higher than the values estimated by
\citet{MR97}. As was, however, discussed by these authors, the
resulting outflow would have a huge Lorentz factor and might not
by itself be able to reproduce the observed properties of real
GRBs.

Although the presentation in this paper and its companion is
focused on the application to GRBs, our formalism is quite
general and the solutions that we present should be
relevant also to other magnetically driven relativistic outflow
sources, notably AGNs and microquasars. The
possible relationship between the different classes of beamed
relativistic outflows in astrophysics has already been noted
before by various authors. One potential attraction of adopting a common modeling
framework for these outflows is that it could shed new light on each
of the different types of sources. For example, the concept of
internal shocks was originally proposed in the context of
AGN jets \citep{R78}, and it has recently been adopted also for
modeling microquasars \citep{KSS00}. In the latter class of
objects, there is direct observational evidence (in the form of
correlated X-ray, infrared, and radio flux variations and the appearance of
superluminal radio knots) for a causal connection between a
rapid accretion episode onto the central black hole and the
ejection of superluminal blobs \citep[e.g.,][]{MR99}. This is the same basic scenario
as the one commonly adopted (albeit without direct observational support)
for GRB outflows. A similar correlation between the X-ray and
radio flux behavior and the appearance of radio-bright superluminal knots was
recently discovered also in an AGN jet \citep{Ma02}. Although the
observed behavior in this case does not support the notion that
AGNs are a simple scaled-up version
of microquasars, this might be attributable to a different
origin of the underlying disk instability that induces the
accretion/ejection episodes \citep[e.g.,][]{MQ01} or to different
environmental conditions \citep[e.g.,][]{H02}. 

There is now also
evidence that the high-energy (GeV-range) $\gamma$-ray emission
in the blazar class of AGNs originates in superluminal radio
knots --- which likely represent discrete blobs or shocks in a
relativistic jet --- at a large distance from the
origin \citep[e.g.,][]{J01b}.\footnote{Although the
bulk Lorentz factors inferred from the apparent superluminal
motions in blazars are not as high as those indicated in GRB
outflows, they can nevertheless be substantial, possibly
exceeding 40 in a number of cases \citep[e.g.,][]{J01a}.}
This is essentially the picture envisioned for GRB
outflows. The superluminal knots in blazar jets such as 3C 345
\citep[e.g.,][]{U97} and 3C 279 \citep[e.g.,][]{W01} are inferred to
move along distinct curved paths, although in the former case it
has been argued \citep{W96} that at least two of the knots
follow each other along the same trajectory. This behavior is
consistent with the ``patchy shell'' model discussed above in
connection with GRBs. There are also indications that the outflows continue to
accelerate on large scales: for example, the Lorentz factor of
knot C7 in the quasar 3C 345 was inferred to increase from $\sim 3$ to
$\gtrsim 10$ as it moved from a distance of $r\simeq 3\, {\rm
pc}$ from the galactic nucleus to $r \simeq 20\, {\rm pc}$ \citep{U97}. A
scaled-down version of this behavior was found in the radio
galaxy NGC 6251, where knots in the radio jets were inferred to
accelerate from $\sim 0.13\, c$ at $r\approx 0.53\, {\rm pc}$ to
$\sim 0.42\, c$ at $r\approx 1.0\, {\rm pc}$ \citep{S00}. 
Such large-scale acceleration is most
naturally interpreted in terms of a Poynting-dominated jet model
of the type discussed in this paper (N. Vlahakis \& A. K\"onigl,
in preparation).

In conclusion, we have derived in this paper the general
formalism for radially self-similar, relativistic MHD outflows
and presented exact solutions (obtained by solving the Euler and
transfield equations and imposing regularity conditions at the relevant
critical points) to illustrate their basic properties. We
focused on trans-Alfv\'enic flows, deferring a discussion of
super-Alfv\'enic solutions to the companion paper.  We
considered ``hot'' and ``cold'' rapid-rotator flows (the latter
reproducing previous results by \citealt{LCB92} and
\citealt{C94}) as well as slow-rotator flows in the
current-carrying regime (which should apply near
the rotation axis), but also described a ``hot,'' fast-rotator
solution in the return-current regime. In all cases, we
demonstrated that the Poynting flux injected at the source can
be transformed with high ($\sim 50\%$) efficiency into
kinetic energy of relativistic baryons at a large (but finite)
distance from the origin. We concentrated on applications to GRB
outflows, although we pointed out that similar solutions may
describe jet acceleration in AGNs and microquasars. In the
application to GRBs, we presented a particular ``hot,'' fast-rotator 
outflow as a fiducial solution. In this case the outflow is
initially accelerated thermally, with the magnetic field acting only 
to guide and collimate it, but subsequently magnetic effects
become fully dominant and are responsible for the bulk of the
acceleration (which occurs between the classical
fast-magnetosonic and modified fast-magnetosonic surfaces) and for
the final collimation to cylindrical asymptotics. We stress,
however, that this solution is only illustrative and is not
meant to provide an accurate description of a typical GRB
flow.\footnote{For one thing, the value of $\gamma_\infty$ in
this solution is about an order of magnitude higher than the
minimum Lorentz factors typically inferred in real GRBs \citep[e.g.,][]{LS01}.}
As we discuss in Paper II, super-Alfv\'enic outflows 
can also provide plausible representations of GRB outflows;
their basic properties, however, are quite similar to those of
trans-Alfv\'enic flows. We have, furthermore,
concluded that solutions in the return-current regime may also be
relevant to these flows; we address this possibility in greater
detail in a separate publication \citep{VPK03}.

\acknowledgements 
This work was supported in part by NASA grant
NAG5-12635 and by the U.S. Department of Energy under grant B341495
to the Center for Astrophysical Thermonuclear Flashes at the University of Chicago.
N. V. also acknowledges support from a McCormick Fellowship at
the Enrico Fermi Institute.

\appendix

\section{Equations in the Axisymmetric Case with E$_{\phi}=0$}\label{RMHDeqs}

We may combine equations (\ref{int-Omega}), (\ref{int-L}), and (\ref{int-mu}) to obtain
$\gamma\,, V_{\phi}$, and $B_{\phi}$ as functions of $M\,,$ $x$, and $G$ using
\begin{equation}\label{A1}
{\boldsymbol {B}} = \frac{\nablas A \times \hat{\phi} }{\varpi} 
-\frac{\mu c \Psi_A x_A^2}{x} \frac{1-G^2}{1-M^2-x^2} \hat
{\phi}\,, \quad
{\boldsymbol{E}}=-\frac{\Omega}{c} \nablas A\,,
\end{equation}
\begin{equation}\label{A2}
{\boldsymbol{V}}=\frac{\Psi_A}{4 \pi \gamma \rho_0} {\boldsymbol{B}} + \varpi \Omega \hat{\phi}
=\frac{\Psi_A}{4 \pi \gamma \rho_0} 
\frac{\nablas A \times \hat{\phi} }{\varpi}
+\frac{\varpi_A \Omega }{G} \frac{G^2-M^2-x^2}{1-M^2-x_A^2}\hat{\phi}
 \,, \quad
\gamma=\frac{\mu}{\xi} \frac{1-M^2-x_A^2}{1-M^2-x^2}\,,
\end{equation}
\begin{equation}\label{rho_P}
\rho_0=\frac{\xi \Psi_A^2}{4 \pi M^2} \,, \quad
P=Q \rho_0^{\Gamma} \,, \quad
\xi= 
1 + \left(\int_0^{P} \frac{dP}{\rho_0 c^2}\right)_{\{s\,, \ A\,
=\, const \}} =
1+\frac{\Gamma}{\Gamma-1}\frac{P}{\rho_0 c^2} \,.
\end{equation}

Knowing the field-line constants (for each $s$)
\begin{equation}
\varpi_A(A\,, s)\equiv \left (\frac{L}{\mu \Omega}\right )^{1/2}\,, 
\quad
x_A(A\,, s) \equiv \frac{\varpi_A \Omega }{c} \,,
\quad
\mu(A\,, s)\,, 
\quad
\sigma_M(A\,, s) \equiv \frac{A \Omega^2 }{ c^3 \Psi_A} \,,
\quad
q(A\,, s) \equiv \frac{\Psi_A^2}{4 \pi}
\left(\frac{\Gamma}{\Gamma-1} \frac{Q}{c^2}\right)^{\frac{1}{\Gamma-1}}
\,,
\end{equation}
we can find the quantities $x\,, M\,, G \,, \xi$ (which in general are functions
of $A\,, \ell$, and $s$) by solving the following system of equations:

\begin{equation}\label{x-M}
x=x_A G \,, \quad 
M^2=q \frac{\xi}{\left(\xi-1\right)^{\frac{1}{\Gamma-1}}}\ ,
\end{equation}
the Bernoulli equation
\begin{equation}\label{bernoulli}
\frac{\mu^2}{\xi^2} \frac{G^2 (1-M^2-x_A^2)^2-x_A^2 (G^2-M^2-x^2)^2}{G^2 (1-M^2-x^2)^2}
=1 + \frac{\sigma_M^2}{\xi^2} \frac{M^4}{x^4} \left(\frac{\varpi \nablas A}{A}
\right)^2 \ ,
\end{equation}
which is obtained after substituting all quantities in the identity
$\gamma^2 \left(1-{V_{\phi}^2}/{c^2}\right) =1+\gamma^2 {V_p^2}/{c^2}$ 
using equations (\ref{A1},\ref{A2}) 
and which in the nonrelativistic limit takes the familiar form
(after Taylor expanding in $1/c^2$)
$$\displaystyle{ \frac{V^2}{2}+\frac{\Gamma}{\Gamma-1} \frac{P}{\rho_0} -
\frac{\varpi \Omega B_{\phi}}{\Psi_A}= (\mu -1)c^2}\,, $$
and the transfield equation (obtained from the component of the
momentum equation along $-{\boldsymbol{\nabla}}A$)
\begin{eqnarray}\label{transfield}
&&
\left[
x^2 \left(\nablas A \right)^2 
\frac{\partial \ln \left(\displaystyle \frac{x_A(A\,,s)}{\varpi_A(A\,,s)} \right)}{\partial A}
-\L A (1-M^2-x^2) \right] \left(\frac{\nablas A}{\varpi} \right)^2 +
\nonumber \\ && +
\left[
\frac{2 x_A^2}{\varpi_A^3 G} \left(\nablas A \right)^2+
\frac{\mu^2 x_A^6 A^2}{\varpi_A^5 \sigma_M^2 M^2 G^3}
\left(\frac{G^2-M^2-x^2}{1-M^2-x^2}\right)^2\right]
\hat{\varpi} \cdot \nablas A -
\nonumber
\\
&&
-\frac{M^2}{2} \nablas
\left[\left(\frac{\nablas A }{\varpi} \right)^2\right]
\cdot \nablas A
-\frac{\Gamma-1}{\Gamma} \nablas \left[
\frac{\xi (\xi-1)}{M^2} 
\frac{A^2 x_A^4}{\sigma_M^2 \varpi_A^4}
\right] \cdot \nablas A -
\nonumber \\ && - 
\frac{1}{2 \varpi^2} \nablas \left[
\frac{\mu^2 A^2 x_A^6}{\sigma_M^2 \varpi_A^2}
\left(\frac{1-G^2}{1-M^2-x^2}\right)^2\right] \cdot \nablas A =0\,,
\end{eqnarray}
where the operator
$\L \equiv {\boldsymbol{\nabla}}^2 - (2/ \varpi){\hat{\varpi}} \cdot {\boldsymbol{\nabla}}$
is related to the curvature radius
${\cal R}= \mid {\boldsymbol{\nabla}} A \mid \left( 
\L A - {\boldsymbol{\nabla}} A \cdot {\boldsymbol{\nabla}} 
\ln \mid {\boldsymbol{\nabla}} A / \varpi \mid \right)^{-1}$.

The latter equation can be written as
\begin{mathletters}\label{forces-perp}
\begin{equation}
{\mbffone}_{G \bot} + {\mbffone}_{T \bot} +
{\mbffone}_{C \bot} + {\mbffone}_{I \bot} + {\mbffone}_{P \bot}
+{\mbffone}_{E \bot}+{\mbffone}_{B \bot}=0 \,,
\end{equation}
where the subscript $\bot$ denotes the component of a force
perpendicular to the poloidal fieldline and pointing toward the axis
(along $\hat{n}= \hat{b} \times \hat{\phi} =
-{\boldsymbol{\nabla}} A/ \mid {\boldsymbol{\nabla}} A \mid$),
and where the individual terms are given by
\begin{eqnarray}
&& {\mbffone}_{G \bot}=0\, , \\ 
&& {\mbffone}_{T \bot}=0\, ,\\ 
&& {\mbffone}_{C \bot}=
-\frac{\gamma^2 \rho_0 \xi V_{\phi}^2}{\varpi} \sin \psi = 
-\frac{B_0^2}{4 \pi \varpi_0}
\frac{c^2 \Psi_A^2}{B_0^2} \frac{x_A^2 \mu^2}{\alpha^{1/2}} \frac{1}{M^2 G^3}
\left(\frac{G^2-M^2-x^2}{1-M^2-x^2}\right)^2 
\frac{\hat{\varpi} \cdot \nablas \alpha}{\mid
\nablas \alpha \mid}\ ,
\label{fcbot}
\\ &&
{\mbffone}_{I \bot}= -\frac{\gamma^2 \rho_0 \xi V_p^2}{{\cal R}}  =
-\frac{M^2 B_p^2}{4 \pi {\cal R}}=
\frac{B_0^2}{4 \pi \varpi_0} \left \{
\frac{M^2 \varpi_0^5}{8}
\nablas \left[\left(\frac{
\nablas {\cal A}}{\varpi} \right)^2\right]
\cdot \frac{\nablas \alpha}{\mid \nablas \alpha \mid}-
\frac{ M^2 \varpi_0^3 \mid \nablas {\cal
A}\mid}{4 \alpha G^2} \L {\cal A} \right\}\ ,
\label{fibot}
\\ &&
{\mbffone}_{P \bot}=\frac{B_0^2}{4 \pi \varpi_0}
\frac{\Gamma-1}{\Gamma} \varpi_0 \nablas \left[
\frac{\xi (\xi-1)}{M^2} \frac{c^2 \Psi_A^2}{B_0^2}\right] \cdot
\frac{\nablas \alpha}{\mid
\nablas \alpha \mid}\ ,
\\ &&
{\mbffone}_{E \bot}=-\frac{B_0^2}{4 \pi \varpi_0}
\frac{x_A^2 \varpi_0 \mid \nablas {\cal A} \mid}{4 \alpha}
\left[\varpi_0^2 \L {\cal A}+\frac{2 }{G \alpha^{1/2}} \varpi_0
\hat{\varpi} \cdot \nablas {\cal A}
+ \left(\varpi_0 \nablas {\cal A} \right)^2
\left(\frac{1}{x_A}\frac{dx_A}{{\cal A}}-\frac{1}{2
\alpha}\frac{d \alpha}{d {\cal A}}\right)\right]\ ,
\\ &&
{\mbffone}_{B \bot}=\frac{B_0^2}{4 \pi \varpi_0} \left \{
\frac{1}{2 \alpha G^2} \varpi_0 \nablas
\left[ \alpha x_A^2 \mu^2 \frac{c^2 \Psi_A^2}{B_0^2}
\left(\frac{1-G^2}{1-M^2-x^2}\right)^2\right] \cdot
\frac{\nablas \alpha}{\mid \nablas \alpha \mid}
+\frac{\varpi_0 \mid \nablas {\cal A} \mid}{4
\alpha G^2} \varpi_0^2 \L {\cal A} \right\}\ .
\end{eqnarray}
\end{mathletters}

\section{Equations in the $r$ Self-Similar Case}\label{appendixrss}

In the $r$ self-similar model described in \S \ref{r-ss_section},
the integrals of motion have the following form:
\begin{mathletters}\label{integrals}
\begin{equation}
\sigma_M= \sigma_0 \left[1-\left(\frac{\alpha_0}{\alpha}\right)^{\frac{F}{2}}\right]\,,
\quad \mbox{where} \quad \sigma_0\,, \alpha_0=const\,,
\quad \mu\,, x_A\,, q=const\,, \quad 
A=\frac{B_0 \varpi_0^2}{F}
\left(\alpha^{\frac{F}{2}}- \alpha^{\frac{F}{2}}_0 \right)\ ,
\end{equation}
with the fieldline constants given by
\begin{equation}
\Psi_A=\frac{B_0 x_A^2}{F \sigma_0 c } \alpha^{\frac{F-2}{2}} \,, \quad
\Omega=\frac{x_A c}{\varpi_0 \alpha^{1/2}} \,, \quad
L=c \varpi_0 x_A \mu \alpha^{1/2} \,, \quad
Q=c^2 \frac{\Gamma-1}{\Gamma}\left(
\frac{4 \pi c^2 F^2 \sigma_M^2 q }{x_A^4 B_0^2 \alpha^{F-2}}
\right)^{\Gamma-1} \,.
\end{equation}
Note that $\sigma_M$ is the ``magnetization parameter'' in the monopole solution of
\citet{M69}.
\end{mathletters}

In cases where $F>0$, the requirement that the magnetic flux function
$A$ vanishes on the rotation axis $\alpha=0$
implies that $\alpha_0=0$ when $\sigma_M=const$ and
$A=\frac{B_0 \varpi_0^2}{F} \alpha^{\frac{F}{2}}$.
We examine solutions with $\alpha_0=0$ throughout this paper.

\begin{mathletters}\label{theta-eqs}
Equations (\ref{x-M}) and (\ref{bernoulli}) take the form
\begin{equation}\label{x-rss}
x=x_A G \,,
\end{equation}
\begin{equation}\label{M-rss}
M^2=q \frac{\xi}{\left(\xi-1\right)^{\frac{1}{\Gamma-1}}}\ ,
\end{equation}
\begin{equation}\label{bernoulli-rss}
\frac{\mu^2}{\xi^2}\frac{G^4 \left(1-M^2-x_A^2\right)^2-
x^2 \left(G^2-M^2-x^2\right)^2}{G^4 \left(1-M^2-x^2\right)^2}=1+
\frac{F^2 \sigma_M^2 M^4 \sin^2 \theta}{\xi^2 x^4 \cos^2
\left(\psi+\theta\right)}\ ,
\end{equation}
and can be solved for $x\,, \xi$, and $\psi$.

Using $\ \tan \psi = \left( \partial z / \partial \varpi
\right)_A$ and equation (\ref{alpha}), we get
$\ \tan \psi = d(G/ \tan \theta) / d G\, ,$ or,
\begin{equation}\label{G-rss}
\frac{d G^2}{d \theta}=\frac{2 G^2 \cos \psi}
{\sin \theta \, \cos \left(\psi+\theta\right)}\ .
\end{equation}

The transfield equation (\ref{transfield}) becomes
\begin{eqnarray}\label{transfield-rss}
G \sin^2 \theta \frac{d}{d\theta}\left[\tan\left(\psi+\theta\right) \frac{1-M^2-x^2}{G} \right]
=&&
\left(F-1\right) \frac{x_A^4 \mu^2 x^2}{F^2 \sigma_M^2 }
\left( \frac{1-G^2}{1-M^2-x^2} \right)^2
-\sin^2 \theta \frac{M^2+F x^2 -F +1}{\cos^2 \left(\psi+\theta\right)}-
\nonumber
\\
&&
-\frac{x_A^4 \mu^2 x^2}{F^2 \sigma_M^2 M^2 }
\left( \frac{G^2-M^2-x^2}{1-M^2-x^2} \right)^2
+2 \frac{\Gamma-1}{\Gamma}
\frac{F-2}{F^2 \sigma_M^2} \frac{\xi \left(\xi-1\right)x^4}{M^2 }\ .
\end{eqnarray}
\end{mathletters}

The magnetization function $\sigma$ is defined as 
the Poynting-to-matter energy flux ratio:
\begin{equation}\label{sigma}
\sigma = \frac{-\varpi \Omega B_\phi / \Psi_A c^2}{\xi \gamma} = 
\frac{x_A^2 - x^2}{1-M^2-x_A^2}\,.
\end{equation}

At the Alfv\'en singular point $\theta =\theta_A$, $G=1$,
$x=x_A$, $M^2=1-x_A^2$, $\xi=\xi_A$, $\psi=\psi_A$, $\sigma = \sigma_A$,
and, using l'H${\hat {\rm o}}$spital's rule,
\begin{eqnarray}\label{reg_A}
\sigma_A = \frac{2 x_A^2 \cos \psi_A}{p_A \sin \theta_A \cos (\theta_A + \psi_A)}\,,
\nonumber \\
\left[\frac{1-M^2-x_A^2}{1-M^2-x^2}\right]_A=\frac{1}{\sigma_A+1}\,,
\
\left[\frac{1-G^2}{1-M^2-x^2}\right]_A =\frac{\sigma_A/x_A^2}{\sigma_A+1} \,,
\
\left[\frac{G^2-M^2-x^2}{1-M^2-x^2}\right]_A =\frac{x_A^2-(1-x_A^2)\sigma_A}{x_A^2 (\sigma_A+1)} \ .
\end{eqnarray}
Substituting the above ratios into equation
(\ref{bernoulli-rss}) yields 
$\mu$ as a function of $\psi_A$, $\theta_A$, and $\sigma_A$,
\begin{equation}\label{alfv1}
\mu^2 = \frac{(\sigma_A+1)^2}{x_A^2-\left[x_A^2-\sigma_A (1-x_A^2) \right]^2}
\left[x_A^2 \xi_A^2 + \frac{F^2 \sigma_M^2 (1-x_A^2)^2 \sin^2 \theta_A}
{x_A^2 \cos^2 (\theta_A + \psi_A)}\right]\,,
\end{equation} 
whereas equation (\ref{transfield-rss})
gives the Alfv\'en regularity condition
\begin{eqnarray}\label{alfv2}
&&
\frac{F^2 \sigma_M^2 (1-x_A^2) (\sigma_A+1)^2 \sin \theta_A }{\mu^2 \cos^2 (\theta_A + \psi_A)}
\nonumber \\ &&
\left\{-2 \frac{\Gamma-1}{\Gamma} \frac{(F-2) (\xi_A-1) (1-x_A^2)}{\xi_A x_A^2} \sin \theta_A
+2 \cos \psi_A \sin (\theta_A + \psi_A) \frac{\sigma_A+1}{\sigma_A}
+\frac{\sin \theta_A }{ x_A^2} \left[ (F-1) (1-x_A^2) -1 \right] \right\}
\nonumber
\\
&&
=
\left[x_A^2-\sigma_A (1-x_A^2) \right]^2 - (F-1)\sigma_A^2 (1-x_A^2) -
2 \frac{\Gamma-1}{\Gamma}(F-2)\frac{\xi_A-1}{\xi_A} 
\left\{x_A^2-\left[x_A^2-\sigma_A (1-x_A^2) \right]^2 \right\} 
\ .
\end{eqnarray} 

Next we prove the following statement:
if one chooses $\Gamma\,, F$ (the model parameters) and
specifies seven boundary conditions on a cone $\theta=const$,
then it is possible to derive all the other model parameters,
and the solution is uniquely determined.

\noindent
{\em Proof:} Suppose that the quantities
$B_{\theta}$, $B_{\phi}$, $V_r$, $V_{\theta}$, $V_{\phi}$, $\rho_0$, and $P$
are given as functions of $r$ along the cone $\theta=const$; i.e.,
$B_{\theta}=-{\cal C}_1 r^{F-2}$, $B_{\phi}=-{\cal C}_2 r^{F-2}$, $V_r/c={\cal C}_3 $,
$V_{\theta}/c=-{\cal C}_4$, $V_{\phi}/c={\cal C}_5$, $\rho_0={\cal C}_6 r^{2(F-2)}$, and
$P={\cal C}_7 r^{2(F-2)}$.
Then one finds
\begin{mathletters}\label{invert}
\begin{eqnarray}
&&
\gamma=1/(1-V^2/c^2)^{1/2}
\,, \quad 
x=(V_{\phi}-V_{\theta} B_{\phi}/B_{\theta})/c 
\,, \quad
\psi=\pi - \theta - \arctan (-V_r / V_{\theta})
\,, \quad 
\xi = 1+ \frac{\Gamma}{\Gamma-1} \frac{P}{\rho_0 c^2}
\,, \quad
\\ &&
M^2=4 \pi \rho_0 \xi (\gamma V_{\theta}/B_{\theta})^2
\,, \quad 
q=M^2 (\xi-1)^{\frac{1}{\Gamma-1}} / \xi
\,, \quad 
\sigma_M = -\gamma \xi x^2 V_{\theta} / c F M^2 \sin \theta
\,, \quad 
\mu = \xi \gamma - \frac{x B_{\phi} /B_{\theta}}{4 \pi \gamma \rho_0 c V_{\theta}}
\,, \quad
\\ &&
x_A^2= \frac{x^2 - (1-M^2) x V_\phi/c}{M^2 + x^2 - x V_\phi/c}
\,, \quad
G=x/x_A
\,, \quad
\mbox{and} \quad 
B_0 \varpi_0^{2-F}= - r^{2-F} B_{\theta}  G^F / (\sin\theta)^{F-1}
\ .
\end{eqnarray}
\end{mathletters}
\noindent
One is thus in a position to start the integration, using
equations (\ref{bernoulli-rss}), (\ref{G-rss}), and (\ref{transfield-rss}), and find a solution
that uniquely corresponds to the seven boundary conditions (or,
equivalently, to the parameters ${\cal C}_j\,,j=1,\cdots,7$) and
the model parameters $\Gamma$ and $F$.
\noindent
{\em QED}.

Whether the solution
actually hits and passes smoothly through any given singular point depends
on the choice of the boundary conditions. From a physical
standpoint, the most robust solutions cross all the three
singular points (the modified slow-magnetosonic, the Alfv\'en,
and the modified fast-magnetosonic points). In this case,
the three regularity conditions at the (a priori unknown) positions of the singular points
impose three relationships among the boundary conditions.

\section{Alfv\'en and Magnetosonic Waves in Relativistic MHD}\label{appendix_waves}

Suppose that we have obtained a solution of the axisymmetric,
relativistic, ideal-MHD equations (\ref{maxwell})--(\ref{energy}). 
If we consider localized,
fast varying, axisymmetric disturbances, then we may assume that
the unperturbed solution is 
uniform and time-independent and neglect all its space and time derivatives.
We may then look for perturbations having a Fourier dependence
$\exp [i(\omega t - {\boldsymbol {k}}\cdot {\boldsymbol {r}})] = 
\exp [i(\omega_{\rm co} t_{\rm co} - {\boldsymbol {k}}_{\rm
co}\cdot {\boldsymbol {r}}_{\rm co})]$,
where, by using the Lorentz transformations between the comoving
$({\boldsymbol {r}}_{\rm co}\,,t_{\rm co})$
and observer's frame $({\boldsymbol {r}}\,,t)$, we get
$\omega_{\rm co}=\gamma \left(\omega - {\boldsymbol {k}}\cdot {\boldsymbol {V}}\right)\,,
{\boldsymbol {k}}_{\rm co} = {\boldsymbol {k}} - {\boldsymbol {V}} \left[
{\gamma \omega }/{c^2} - \left(\gamma -1 \right) {{\boldsymbol
{k}}\cdot {\boldsymbol {V}}}/{V^2}\right]$.\footnote{
Note that, for axisymmetric disturbances in the observer's
frame, the wavevector ${\boldsymbol {k}}$ lies in the
meridional plane,
but that, nonetheless, the comoving wavevector ${\boldsymbol
{k}}_{\rm co}$ has an azimuthal component!}

It is more convenient to analyze the disturbances in the (comoving) flow frame.
Define a local Cartesian system of coordinates $(x\,,y\,,z)$ such that
${\boldsymbol {B}}_{\rm co}= B_{\rm co} \hat{z}$, and 
${\boldsymbol {k}}_{\rm co}=k_{\rm co}(\hat{z} \cos \theta_0 +
\hat{x} \sin \theta_0)$. After linearizing equations
(\ref{maxwell})--(\ref{energy}), we may express all the perturbed quantities
in terms of the perturbation $\delta {\boldsymbol {V}}_{\rm co}$.
After some manipulation, we obtain
\begin{eqnarray}
\displaystyle
&&
\left(\begin{array}{ccc}
D_{11} & 0 & D_{13} \\
0 & D_{22} & 0 \\
D_{13} & 0 & D_{33}
\end{array}\right)
\left(\begin{array}{c}
\hat{x} \cdot \delta {\boldsymbol {V}}_{\rm co} \\
\hat{y} \cdot \delta {\boldsymbol {V}}_{\rm co} \\
\hat{z} \cdot \delta {\boldsymbol {V}}_{\rm co} 
\end{array}\right)
 = 0\,, \quad \mbox{where}
\nonumber  \\
&&
D_{11}= 
\frac{c_s^2}{c^2} \sin^2 \theta_0 +
\left(\frac{v_A^2}{c^2}-\frac{\omega_{\rm co}^2}{c^2 k_{\rm co}^2}\right)
\left(1-\frac{v_A^2}{c^2}\right)^{-1} \,,
\quad
D_{13}=\frac{c_s^2}{c^2} \sin \theta_0 \cos \theta_0 \,,
\nonumber \\
&&
D_{22}=\left(\frac{v_A^2}{c^2} \cos ^2 \theta_0 -
\frac{\omega_{\rm co}^2}{c^2 k_{\rm co}^2}\right)
\left(1-\frac{v_A^2}{c^2}\right)^{-1} 
\,,
\quad 
D_{33}=\frac{c_s^2}{c^2} \cos^2 \theta_0-
\frac{\omega_{\rm co}^2}{c^2 k_{\rm co}^2} \ ,
\nonumber 
\end{eqnarray}
where the Alfv\'en speed can be expressed in terms of the
corresponding proper speed $U_A$, which satisfies
$U_A^2\equiv v_A^2/(1-v_A^2/c^2) = B_{\rm co}^2/4 \pi \rho_0 \xi$.

Besides the trivial entropy wave $\omega_{\rm co}=0$ (which, however, corresponds to 
$\omega \neq 0$!),
the dispersion relation $\mid D \mid = 0$ yields the wave modes
listed below.
\begin{itemize}
\item
Alfv\'en waves: $D_{22}=0$, or $\omega_{\rm co} / k_{\rm co} = \pm v_A \cos \theta_0
$, corresponding to a displacement $\delta {\boldsymbol {V}}_{\rm co} $ normal to the
\{${\boldsymbol {B}}_{\rm co} \,, {\boldsymbol {k}}_{\rm co}$\} plane.
Transforming the dispersion relation to the observer's frame, we get
\begin{equation}\label{waves_alfven}
\left(\gamma \ \frac{\omega - {\boldsymbol {k}} \cdot {\boldsymbol {V}}}{c k}\right)^2=
\frac{\left({\boldsymbol {B}} \cdot {\boldsymbol {k}} / k\right)^2}{4 \pi \rho_0 \xi c^2}
\left[1- \left(x+\frac{\omega B_{\phi}}{c {\boldsymbol {B}} \cdot {\boldsymbol {k}}} \right)^2 
- \left(\frac{\omega B_p}{c {\boldsymbol {B}} \cdot {\boldsymbol
{k}}}\right)^2 \right]\ .
\end{equation}
\item
Slow/fast-magnetosonic waves:
$D_{11} D_{33}= D_{13}^2$, or,
$\displaystyle{
\left(\frac{\omega_{\rm co}}{ c k_{\rm co}}\right)^4-
\left(\frac{\omega_{\rm co}}{ c k_{\rm co}}\right)^2
\left(\frac{c_s^2}{c^2}+\frac{v_A^2}{c^2}-
\frac{c_s^2 v_A^2}{c^4} \sin^2 \theta_0 \right) + 
\frac{c_s^2 v_A^2}{c^4} \cos^2 \theta_0 = 0
}
$, corresponding to a displacement 
$\delta {\boldsymbol {V}}_{\rm co} $ in the \{${\boldsymbol
{B}}_{\rm co} \,, {\boldsymbol {k}}_{\rm co}$\} plane.
In the observer's frame, we have
\begin{eqnarray}\label{waves_fast}
\left(1-\frac{\omega^2}{c^2 k^2}\right)^{-1}
\left(\gamma \ \frac{\omega - {\boldsymbol {k}} \cdot {\boldsymbol {V}}}{c k}\right)^4-
\nonumber \\
\left(\gamma \ \frac{\omega - {\boldsymbol {k}} \cdot {\boldsymbol {V}}}{c k}\right)^2
\left(\frac{U_s^2}{c^2}+
\frac{B^2 - E^2}{4 \pi \rho_0 \xi c^2}\right)+
\frac{U_s^2}{c^2}
\frac{\left({\boldsymbol {B}} \cdot {\boldsymbol {k}} / k\right)^2}{4 \pi \rho_0 \xi c^2}
\left[1- \left(x+\frac{\omega B_{\phi}}{c {\boldsymbol {B}} \cdot {\boldsymbol {k}}} \right)^2 
- \left(\frac{\omega B_p}{c {\boldsymbol {B}} \cdot {\boldsymbol
{k}}}\right)^2 \right]=0 \ .
\end{eqnarray}
\end{itemize}

An interesting property of the waves, related to the discussion on
critical/singular surfaces of steady-state MHD, is the following:
If the component of the flow proper velocity along the wavevector
equals in magnitude, but is opposite in sign, to the comoving proper phase velocity of the wave,
then $\omega=0$ and the wave is static. (The converse is also true.)
Thus,
\begin{equation}\label{theorem}
\omega =0 \quad \Leftrightarrow \quad
\gamma {\boldsymbol {V}} \cdot \frac{{\boldsymbol {k}}}{k}=
-\frac{\omega_{\rm co}/k_{\rm co}}{\left(1-\omega_{\rm co}^2/c^2 k_{\rm co}^2\right)^{1/2}} \,.
\end{equation}
This statement is easily proved by combining 
$\omega_{\rm co}=\gamma \left(\omega - {\boldsymbol {k}}\cdot {\boldsymbol {V}}\right)$
with the Lorentz invariant $\omega_{\rm co}^2-c^2 k_{\rm co}^2 = \omega^2-c^2 k^2$
and solving for $\omega$.
\\
Equation (\ref{theorem}) is the generalization of the property of nonrelativistic
static waves ${\boldsymbol {V}} \cdot {\boldsymbol {k}} + \omega_{\rm co}  = \omega = 0$.
It is consistent with the result that proper speeds are the appropriate generalization
of ordinary speeds in relativistic theory \citep[e.g.,][]{K80}.

\end{document}